\documentclass[showpacs,aps,prd,reprint,superscriptaddress,nofootinbib]{revtex4-1}
\usepackage[colorlinks=true, pdfstartview=FitV, linkcolor=magenta,citecolor=blue, urlcolor=blue,
bookmarks=true, bookmarksnumbered=true, breaklinks]{hyperref}
\usepackage[dvipdfmx]{graphicx}
\usepackage{amsmath,amssymb,bm,color,longtable,mathrsfs,slashed}
\usepackage{ulem}

\def\be#1{\begin{equation}#1\end{equation}} 

\def\beqnn#1{\begin{eqnarray}#1\end{eqnarray}}
\def\pdif#1#2{\frac{\partial {#1}}{\partial {#2}}} 

\begin{document}
\begin{flushright}
KEK-TH-2097
\end{flushright}

\title{Casimir effect for nucleon parity doublets}

\author{Tsutomu~Ishikawa}
\email[]{{\tt tsuto@post.kek.jp}}
\affiliation{Graduate University for Advanced Studies (SOKENDAI), Tsukuba, 305-0801, Japan}
\affiliation{KEK Theory Center, Institute of Particle and Nuclear
Studies, High Energy Accelerator Research Organization (KEK), Tsukuba, 305-0801, Japan}

\author{Katsumasa~Nakayama}
\email[]{{\tt katumasa@post.kek.jp}}
\affiliation{Department of Physics, Nagoya University, Nagoya, 464-8602, Japan}
\affiliation{KEK Theory Center, Institute of Particle and Nuclear
Studies, High Energy Accelerator Research Organization (KEK), Tsukuba, 305-0801, Japan}

\author{Kei~Suzuki}
\email[]{{\tt kei.suzuki@kek.jp}}
\affiliation{KEK Theory Center, Institute of Particle and Nuclear
Studies, High Energy Accelerator Research Organization (KEK), Tsukuba, 305-0801, Japan}

\date{\today}

\begin{abstract}
Finite-volume effects for the nucleon chiral partners are studied within the framework of the parity-doublet model.
Our model includes the vacuum energy shift for nucleons, which is the Casimir effect.
We find that for the antiperiodic boundary the finite-volume effect leads to chiral symmetry restoration, and the masses of the nucleon parity doublets degenerate.
For the periodic boundary, the chiral symmetry breaking is enhanced, and the masses of the nucleons also increase.
We also discuss the finite-temperature effect and the dependence on the number of compactified spatial dimensions.
\end{abstract}


\maketitle

\section{Introduction}
Chiral symmetry is a fundamental property of quarks in QCD.
At low temperature and density, chiral symmetry is spontaneously broken by the chiral condensate, which affects the various properties of hadrons, such as masses and decay constants.
On the other hand, at high temperature and/or density, chiral symmetry is restored by medium effects, and the hadronic observables are drastically modified.
In particular, a useful concept to elucidate the relation between chiral symmetry and hadronic observables is the {\it chiral-partner structure} between hadrons.
This structure means that the masses (or other observables) of the partners split in the chiral-broken phase and become degenerate in the chiral-restored phase.

The parity-doublet model for nucleons was first proposed in Ref.~\cite{Detar:1988kn} to understand a nucleon doublet [e.g., the positive-parity $N$ and negative-parity $N^\ast (1535)$] as a chiral partner.
This model has been applied not only to investigate the role of chiral-partner structures for baryons in vacuum \cite{Detar:1988kn,Nemoto:1998um,Jido:1998av,Jido:1999hd,Jido:2001nt,Chen:2008qv,Gallas:2009qp,Dmitrasinovic:2009vp,Dmitrasinovic:2009vy,Chen:2009sf,Nagata:2010ir,Chen:2010ba,Chen:2011rh,Paeng:2011hy,Gallas:2013ipa,Nishihara:2015fka,Olbrich:2015gln,Dmitrasinovic:2016hup,Olbrich:2017fsd,Catillo:2018cyv,Lakaschus:2018rki,Yamazaki:2018stk} but also to elucidate various physics in nuclear environments such as $\eta$ mesic nuclei \cite{Jido:2002yb,Nagahiro:2003iv,Nagahiro:2005gf,Jido:2008ng,Nagahiro:2008rj}, hadron modification in matter \cite{Kim:1998upa,Harada:2016uca,Suenaga:2017wbb,Suenaga:2018kta}, the phase diagrams of the isospin symmetric nuclear matter \cite{Hatsuda:1988mv,Zschiesche:2006zj,Dexheimer:2007tn,Dexheimer:2008cv,Sasaki:2011ff,Gallas:2011qp,Steinheimer:2011ea,Paeng:2013xya,Heinz:2013hza,Benic:2015pia,Weyrich:2015hha,Motohiro:2015taa,Takeda:2017mrm,Marczenko:2017huu,Takeda:2018ldi,Shin:2018axs}, isospin asymmetric nuclear matter \cite{Dexheimer:2007tn,Dexheimer:2008cv,Motohiro:2015taa,Takeda:2017mrm,Mukherjee:2017jzi}, thermal nuclear matter \cite{Sasaki:2010bp,Steinheimer:2011ea,Mukherjee:2016nhb,Sasaki:2017glk}, and magnetized nuclear matter \cite{Haber:2014ula}, and neutron stars \cite{Hatsuda:1988mv,Dexheimer:2007tn,Dexheimer:2008cv,Dexheimer:2012eu,Schramm:2015hba,Schramm:2015hba,Mukherjee:2017jzi,Marczenko:2018jui,Mukherjee:2018yft}.
Recently, the degeneracy for the correlators (and masses) of the positive- and negative-parity nucleons was found from lattice QCD simulations at high-temperature phase above the chiral phase transition \cite{Aarts:2015mma,Aarts:2017rrl,Aarts:2018glk}.
These results might indicate not only the validity of the parity-doublet picture but also the survival of the {\it chiral invariant mass} (namely, another origin of the nucleon masses) at high temperature.

\begin{figure}[t!]
    \begin{minipage}[t]{1.0\columnwidth}
        \begin{center}
            \includegraphics[clip, width=1.0\columnwidth]{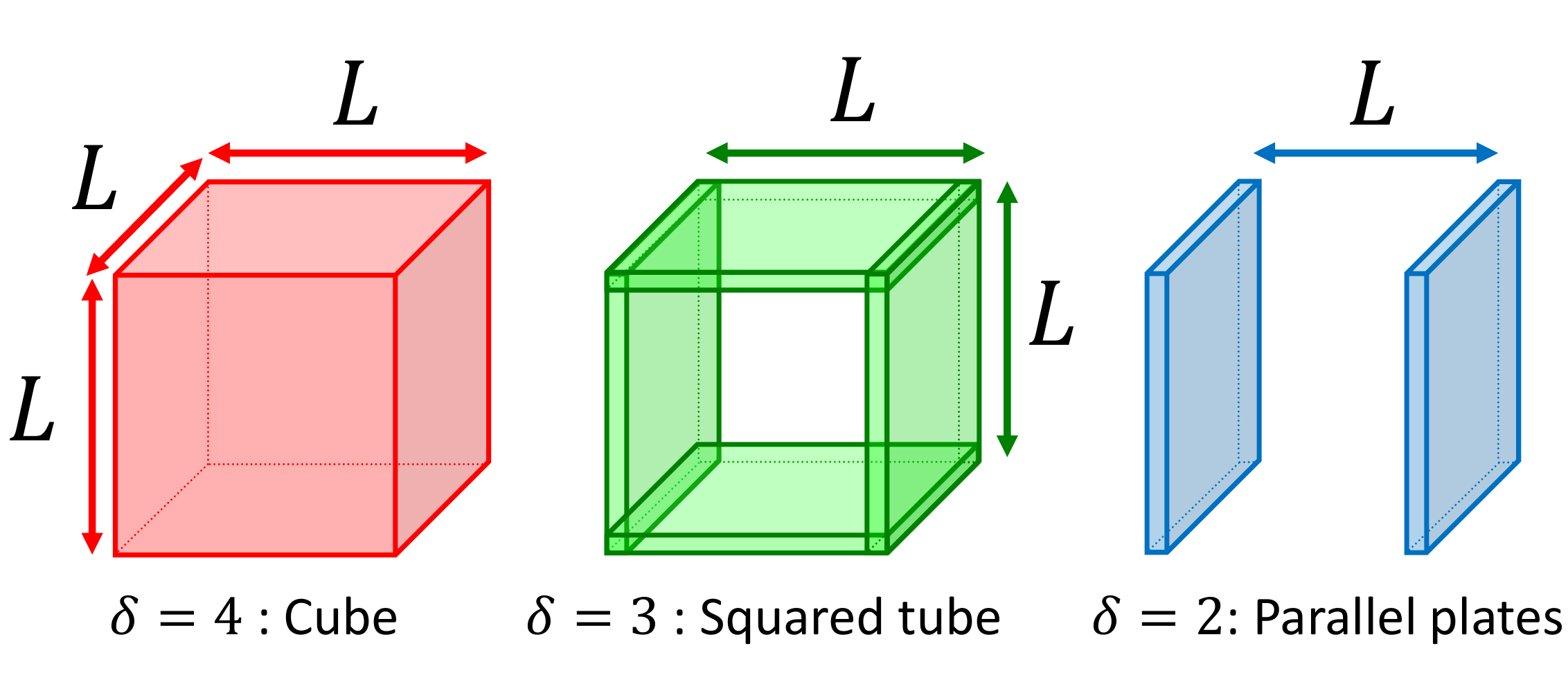}
        \end{center}
    \end{minipage}
    \caption{Box geometry with compactified spatial length $L$ in the $3+1$ dimensional space-time, where $\delta$ is defined as the temporal and compactified spatial dimensions.}
    \label{box}
\vspace{-15pt}
\end{figure}

\renewcommand{\labelenumi}{(\roman{enumi})}

The purpose of this work is to focus on finite-volume effects for the nucleon parity-doublet structure.
Within the parity-doublet model, we consider nucleons inside a finite ``box" with a boundary condition.
Such a setup will enable us to compare results from the models with observables from lattice QCD simulations.
Here, lattice QCD setup has two advantages:
\begin{enumerate}
\item We can compactify the arbitrary space-time dimensions so that we can study not only finite-volume effects in the usual $3+1$ dimensional box but also physics in an ``anisotropic box," such as the (usual) Casimir effect \cite{Casimir:1948dh} induced by one dimensional compactification, as shown in Fig.~\ref{box}.
\item We can choose arbitrary boundary conditions such as periodic and antiperiodic ones, which might modify the infrared part of the momentum of particles.
Thus, our studies will be useful for giving us an intuitive interpretation of the role of chiral symmetry in a finite volume.
\end{enumerate}

\renewcommand{\labelenumi}{(\arabic{enumi})}

It should be noted that finite-volume effects for the nucleon masses in a box could be estimated within the framework of the chiral perturbation theory (ChPT) with baryons \cite{AliKhan:2003ack,Beane:2004tw,Koma:2004wz,Detmold:2004ap,Bedaque:2004dt,Colangelo:2005cg,Colangelo:2010ba,Geng:2011wq},\footnote{For the early works of the finite-volume ChPT (without baryons) by Gasser and Leutwyler, see Refs.~\cite{Gasser:1986vb,Gasser:1987ah,Leutwyler:1987ak,Gasser:1987zq}.} which have been devoted to compare results with artificial volume effects from lattice QCD simulations.
We emphasize that our purpose in this work is to investigate the properties of the nucleon parity doublets in a finite box, in which the finite-volume effects for $\sigma$ mean fields will be essential.
This is a different situation from the ChPT, in which the momentum discretization effects for pion loops would be dominant.

In fact, finite-volume effects for the chiral symmetry breaking/restoration have been investigated by effective models such as the Nambu--Jona-Lasinio model in 3+1 dimensions \cite{Abreu:2006pt,Ebert:2010eq,Hayashi:2010ru,Abreu:2009zz,Abreu:2010zzb,Abreu:2011zzc,Wang:2018ovx,Wang:2018kgj}, 2+1 dimensions \cite{Song:1993da,Kim:1994es,Ebert:2015vua}, and 1+1 dimensions (or the so-called chiral Gross-Neveu model) \cite{Kim:1987db,Song:1990dm,Vshivtsev:1995xx,Vdovichenko:1998ev,Ebert:2008us,Ebert:2011tt,Flachi:2012pf,Flachi:2013bc} and linear sigma (or quark-meson) model \cite{Braun:2004yk,Braun:2005gy,Braun:2005fj,Palhares:2009tf,Braun:2010vd,Braun:2011iz,Tiburzi:2013vza,Tripolt:2013zfa,Phat:2014asa}.
Also, to study the thermodynamics taking into account the deconfinement transition, we can utilize the models implementing the properties of the Polyakov loop, such as the Polyakov-Nambu--Jona-Lasinio model \cite{Bhattacharyya:2012rp,Bhattacharyya:2014uxa,Bhattacharyya:2015kda,Saha:2017xjq} and Polyakov-linear-sigma model \cite{Magdy:2015eda},
The thermodynamics of hadronic matter without quarks could be investigated by the hadron resonance gas model in a finite volume \cite{Bhattacharyya:2015zka,Xu:2016skm,Samanta:2017ohm}.
For nucleon sectors, the Walecka model \cite{Walecka:1974qa} (or sometimes called the $\sigma$-$\omega$ model) is well known as a conventional tool to study properties of the $\sigma$ mean field in the nuclear matter.
Finite-volume effects at zero and finite temperatures from this model were studied in Ref.~\cite{Abreu:2017lxf}.

This paper is organized as follows.
In Sec.~\ref{Sec_2}, we introduce the parity-doublet model in a finite box.
To compare different models, we also review the case of the Walecka model.
Our numerical results are shown in Sec.~\ref{Sec_3}.
Section~\ref{Sec_4} is devoted to our outlook.

\section{Formalism} \label{Sec_2}
In this work, we use two types of models to study the nucleon masses: the Walecka model and parity-doublet model.
After introducing each model, we also introduce the finite-volume effects including the Casimir effects.
Note that, to generalize our formulation, we include the baryon chemical potential in this section, but the numerical results  in Sec.~\ref{Sec_3} are limited to zero chemical potential.

\subsection{Walecka model}
The Lagrangian of the Walecka model \cite{Walecka:1974qa} is
\begin{eqnarray}
\mathcal{L}_\mathrm{Walecka} &=& \bar{\psi} ( i \partial\hspace{-0.55em}/ + \mu_N \gamma_0 - m_N + g_\sigma \sigma - g_\omega \gamma^\mu \omega_\mu )  \psi \nonumber\\
&& + \mathcal{L}_\mathrm{Walecka}^\mathrm{mes},
\end{eqnarray}
where $\psi$ is a nucleon field and $\mu_N$ and $m_N$ are its chemical potential and mass, respectively.
The nucleon field interacts with meson fields by the coupling constants, $g_\sigma$ and $g_\omega$.
For the mesonic part, we include the isoscalar-scalar $\sigma$ and isoscalar-vector $\omega_\mu$ (the field strength tensor is $\omega_{\mu \nu} \equiv \partial_\mu \omega_\nu - \partial_\nu \omega_\mu$):
\begin{equation}
\mathcal{L}_\mathrm{Walecka}^\mathrm{mes} = \frac{1}{2}( \partial_\mu \sigma \partial^\mu \sigma -m_\sigma \sigma^2)  -\frac{1}{4} \omega_{\mu \nu} \omega^{\mu\nu} + \frac{m_\omega^2}{2} \omega_\mu \omega^\mu.
\end{equation}

By the mean-field approximation, we introduce the classical fields $\sigma \to  \bar{\sigma}$ and $\omega^0 \to \bar{\omega}^0$.
The effective nucleon mass $M$ and the effective nucleon chemical potential $\mu^\ast$ are given by
\begin{eqnarray}
M &=& m_N -g_\sigma \bar{\sigma}, \\ 
\mu^\ast &=& \mu_N - g_\omega \bar{\omega}^0.
\end{eqnarray}
The numerical parameters are shown in Table~\ref{Tab_para_Wal}.

\begin{table}[t!]
\centering
\caption{Parameters of the Walecka model \cite{Buballa:1996tm}.}
\begin{tabular}{cc}
\hline\hline
Parameters &  Values \\
\hline
$m_N$ (MeV) & $939$ \\
$m_\sigma$ (MeV) & $550$ \\
$m_\omega$ (MeV) & $783$ \\
$g_\sigma$ & $10.3$ \\
$g_\omega$ & $12.7$ \\
\hline\hline
\end{tabular}
\label{Tab_para_Wal}
\end{table}

\subsection{Parity-doublet model}
The Lagrangian of the parity-doublet model with the mirror assignment \cite{Detar:1988kn} is written as\footnote{In this Lagrangian, the chemical potentials for the bare baryon fields, $\psi_1$ and $\psi_2$, are not defined, but we can introduce the baryon chemical potential $\mu_N$ for the physical nucleon fields after diagonalizing the mass matrix.}
\begin{eqnarray}
\mathcal{L}_\mathrm{Mirror} &=& \bar{\psi}_1 i \partial\hspace{-0.55em}/ \psi_1 + \bar{\psi}_2 i \partial\hspace{-0.55em}/ \psi_2 +m_0 (\bar{\psi}_2 \gamma_5 \psi_1 - \bar{\psi}_1 \gamma_5 \psi_2 ) \nonumber\\
&& + g_1 \bar{\psi}_1 (\sigma + i \gamma_5 \vec{\tau} \cdot \vec{\pi}) \psi_1 + g_2 \bar{\psi}_2 (\sigma - i \gamma_5 \vec{\tau} \cdot \vec{\pi}) \psi_2 \nonumber\\
&& + \mathcal{L}_\mathrm{Mirror}^\mathrm{mes},
\end{eqnarray}
where $\psi_1$ ($\psi_2$) is a ``bare" baryon field with positive (negative) parity and $m_0$ is called the {\it chiral invariant mass} mixing $\psi_1$ and $\psi_2$.
The baryon fields interact with the meson fields by the coupling constants, $g_1$ and $g_2$.
For the mesonic part, we include the isoscalar-scalar $\sigma$, isovector-pseudoscalar $\vec{\pi}$, and isoscalar-vector $\omega_\mu$:
\begin{eqnarray}
\mathcal{L}_\mathrm{Mirror}^\mathrm{mes} &=& \frac{1}{2}( \partial_\mu \sigma \partial^\mu \sigma) + \frac{1}{2}( \partial_\mu \vec{\pi} \partial^\mu \vec{\pi}) \nonumber\\
&& + \frac{\bar{\mu}^2 }{2} (\sigma^2 +\vec{\pi}^2) - \frac{\lambda}{4}  (\sigma^2 +\vec{\pi}^2)^2 \nonumber\\
&& + \epsilon \sigma -\frac{1}{4} \omega_{\mu \nu} \omega^{\mu\nu} + \frac{m_\omega^2}{2} \omega_\mu \omega^\mu, \label{pd_mes}
\end{eqnarray}
where the term with $\epsilon \sigma$ corresponds to the explicitly chiral symmetry breaking.

After applying the mean-field approximation for the scalar and vector fields, $\sigma \to \bar{\sigma}$ and $\omega^0 \to \bar{\omega}^0$, and diagonalizing the mass matrix of the nucleons, we obtain the mass formulas for the nucleon parity doublet:
\begin{equation}
M_{\pm} = \frac{1}{2} \left( \sqrt{(g_1+g_2)^2 \bar{\sigma}^2 +4m_0^2 } \mp (g_1-g_2) \bar{\sigma} \right) \label{mass},
\end{equation}
where $M_+$ and $M_-$ are the ``physical" nucleon masses with the positive and negative parities, respectively.
We note that the $\bar{\sigma}^2$ term of Eq.~(\ref{mass}) lifts up both the masses $M_+$ and $M_-$, while the linear $(g_1-g_2)\bar{\sigma}$ term splits the masses.
As an interesting situation, when $\bar{\sigma}$ is small enough [$\bar{\sigma} \ll 4m_0(g_1-g_2)/(g_1 +g_1)^2$], the linear term contribution dominates the mass formula, so the nucleon masses still split from $m_0$, and the mass of the positive-parity nucleon becomes smaller than $m_0$.
Such a situation will be realized in our numerical results.
The effective baryon chemical potential $\mu^\ast$ is given by
\begin{equation}
\mu^\ast = \mu_N - g_\omega \bar{\omega}^0.
\end{equation}
The numerical parameters based on Ref.~\cite{Zschiesche:2006zj} are shown in Table~\ref{Tab_para_mir}.\footnote{Note that in Ref.~\cite{Zschiesche:2006zj} the parameters were determined to reproduce the properties of the nuclear matter.
Even though these parameters are applied, we could reproduce the physical quantities {\it in vacuum}, such as the decay widths of $N^\ast \to \pi N$, by including additional higher-order derivative coupling terms.}

\begin{table}[t!]
\centering
\caption{Parameters of the parity-doublet model~\cite{Zschiesche:2006zj}, where $f_\pi=93 \mathrm{MeV}$, $m_\pi=138 \mathrm{MeV}$, and $\epsilon =m_\pi^2 f_\pi$.}
\begin{tabular}{cc}
\hline\hline
Parameters &  Values \\
\hline
$m_0$ (MeV) & $790$ \\
$g_1$ & $13.0$ \\
$g_2$ & $6.97$ \\
$g_\omega$ & $6.79$ \\
$\bar{\mu}$ (MeV) & $199.26$ \\
$\lambda$ & $6.82$ \\
\hline\hline
\end{tabular}
\label{Tab_para_mir}
\end{table}

\subsection{Thermodynamic potentials}
The nucleonic part of the thermodynamic potential (per volume $V$) at temperature $T$ is
\begin{eqnarray}
&& \frac{\Omega_N(T,\mu_N)}{V} = - \sum_{i} \gamma_i \int \frac{d^3p}{(2\pi)^3} \left[ E_i(p) \right. \nonumber\\
&& \left. + T \left\{ \ln \left[ 1+e^{-\beta (E_i(p) - \mu_i^\ast)} \right] + \ln \left[1+e^{-\beta (E_i(p) + \mu_i^\ast)} \right] \right\} \right] \label{omega_N},  \nonumber\\
\end{eqnarray}
where the index $i$ of the nucleon degrees of freedom included in the model, labels only $N$ for the Walecka model and $N_+$ and $N_-$ for the parity-doublet model.
$\gamma_i = 2 \times 2$ is the spin-isospin degeneracy factor, and $E_i(p) = \sqrt{p^2+M_i^2}$ is the energy of nucleons.
The first term of Eq.~(\ref{omega_N}) with the ultraviolet divergence corresponds to the free energy of the vacuum, and the second (third) term is the thermal and density effects for nucleons (antinucleons). 

The mesonic parts of the thermodynamic potentials are
\begin{equation}
\frac{\Omega_\mathrm{Walecka}^\mathrm{mes}}{V} = -\frac{1}{2}m_\omega^2 \bar{\omega}_0^2 + \frac{1}{2} m_\sigma^2 \bar{\sigma}^2  \label{omega_meson_Wal},
\end{equation}
\begin{equation}
\frac{\Omega_\mathrm{Mirror}^\mathrm{mes}}{V} = -\frac{1}{2}m_\omega^2 \bar{\omega}_0^2 - \frac{1}{2} \bar{\mu}^2 \bar{\sigma}^2 + \frac{1}{4} \lambda \bar{\sigma}^4  - \epsilon \bar{\sigma} \label{omega_meson_mir}.
\end{equation}
The potential for the whole system is defined by ${\Omega(T,\mu_N)} \equiv \Omega_N(T,\mu_N)+\Omega^\mathrm{mes}(T,\mu_N)$.
We solve the gap equations for the mean fields $\bar{\sigma}$ and $\bar{\omega}_0$, which are represented by $\frac{\partial \Omega(T,\mu_N)}{\partial \bar{\sigma}} = 0$ and $\frac{\partial \Omega(T,\mu_N)}{\partial \bar{\omega}_0} = 0$.

\subsection{Finite-volume effect}
In the following, we introduce the finite-volume effects for the compactified dimension $\delta$ in the $3+1$ dimensional space-time.
Here, we focus on the compactification of the one spatial dimension ($\delta=2$; also see Fig.~\ref{box}), which has the spatial $R^2 \times S^1$ topology.
This setup is the so-called two parallel plates geometry and is the same situation as the original Casimir effect.
For the generalization to arbitrary compactified dimensions, see Appendix~\ref{App_1}.

For $\delta=2$, we discretize the $z$ component of the 3-momentum for nucleon fields:
\begin{eqnarray}
p_z &&\to p_z^\mathrm{ap} = \frac{(2l+1) \pi}{L}, \label{pz_anti} \\
p_z &&\to p_z^\mathrm{p} = \frac{2l \pi}{L}, \label{pz_peri}
\end{eqnarray}
where $l=0, \pm1, \cdots$, for the antiperiodic boundary condition [$\psi(\tau,x,y,z=0) = - \psi(\tau,x,y,z=L)$] and for the periodic boundary condition [$\psi(\tau,x,y,z=0) = \psi(\tau,x,y,z=L)$], respectively.
The resulting energy is represented by $E_i(p) = \sqrt{p_z^2 + p_\perp^2 + M_i^2}$, where $p_\perp^2 =p_x^2+p_y^2$.
The thermodynamic potential is rewritten as
\begin{eqnarray}
&& \frac{\Omega_N(T,\mu_N,L)}{V} = - \sum_{i} \frac{\gamma_i}{L} \sum_{l=-\infty}^\infty \int \frac{d^2 p_\perp}{(2\pi)^2} \left[ E_i(p) \right. \nonumber\\
&& \left. + T \left\{ \ln \left[ 1+e^{-\beta (E_i(p) - \mu_i^\ast)} \right] + \ln \left[1+e^{-\beta (E_i(p) + \mu_i^\ast)} \right] \right\} \right].  \nonumber\\ \label{Omega_V}
\end{eqnarray}
Here, we separate the finite-volume effects into two parts: i) a thermal energy shift for nucleon free energy, which corresponds to the second and third terms of Eq.~(\ref{Omega_V}), and ii) an energy shift for the zero-point energy that is the Casimir effect,\footnote{As shown in Appendix~\ref{App_1}, all the finite-volume effects including the $L$-dependence of the second and third terms as well as the first term could be understood as a kind of Casimir effect.} which corresponds to the first term of Eq.~(\ref{Omega_V}).

\begin{figure}[tb!]
    \begin{minipage}[t]{1.0\columnwidth}
        \begin{center}
            \includegraphics[clip, width=1.0\columnwidth]{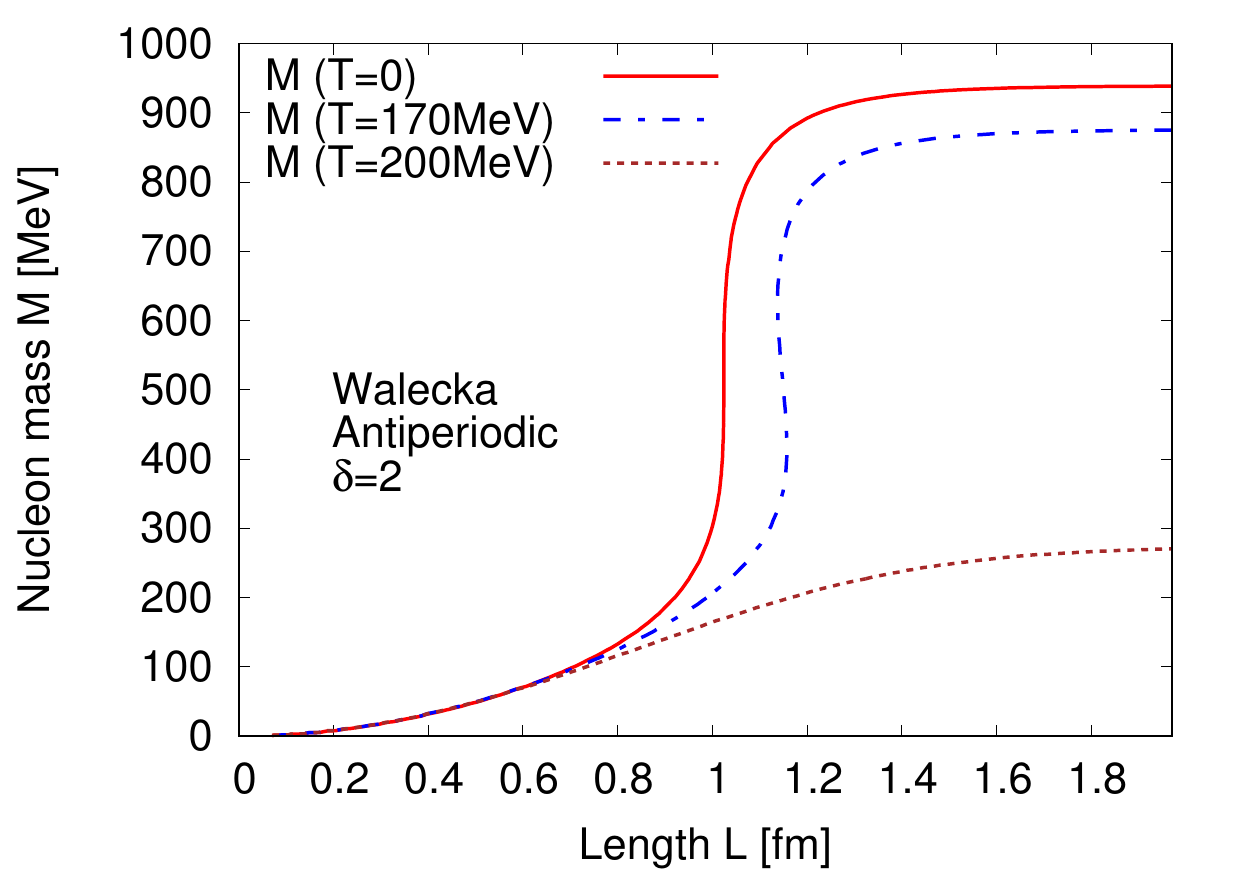}
        \end{center}
    \end{minipage}
    \begin{minipage}[t]{1.0\columnwidth}
        \begin{center}
            \includegraphics[clip, width=1.0\columnwidth]{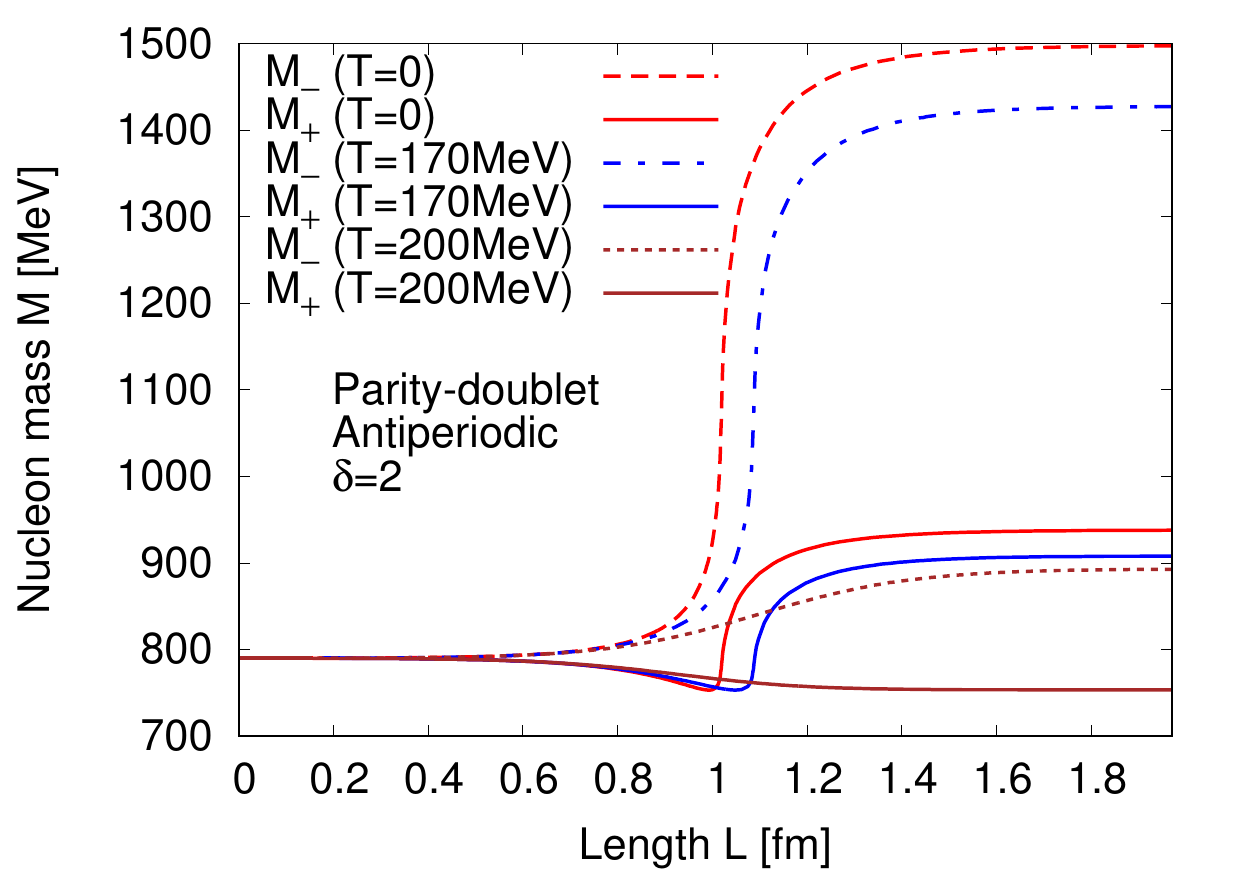}
        \end{center}
    \end{minipage}
    \caption{Finite-volume transition of nucleon masses with $\delta=2$ and antiperiodic boundary condition.
Upper: Walecka model. Lower: Parity-doublet model.}
    \label{apb_LM_delta2}
\end{figure}

\begin{figure}[tb!]
    \begin{minipage}[t]{1.0\columnwidth}
        \begin{center}
            \includegraphics[clip, width=1.0\columnwidth]{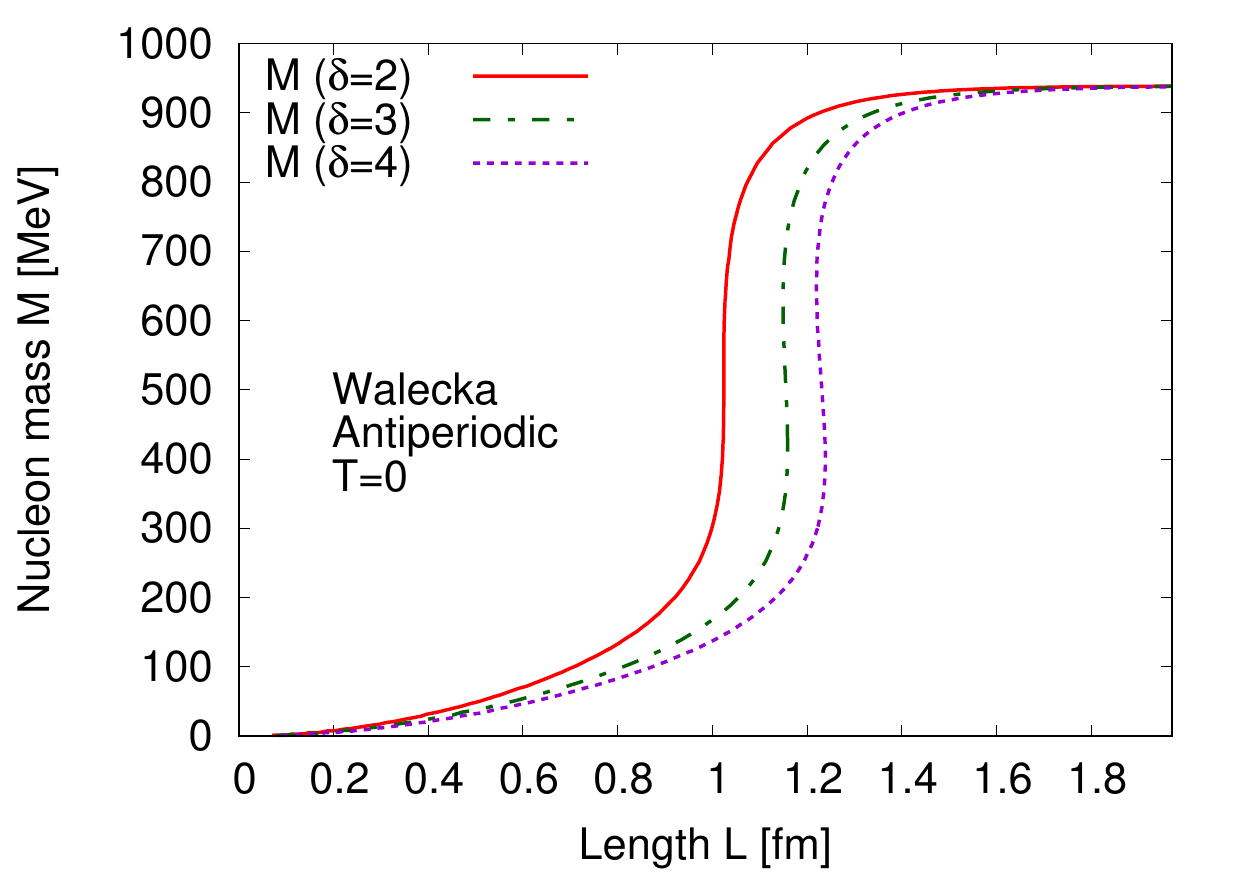}
        \end{center}
    \end{minipage}
    \begin{minipage}[t]{1.0\columnwidth}
        \begin{center}
            \includegraphics[clip, width=1.0\columnwidth]{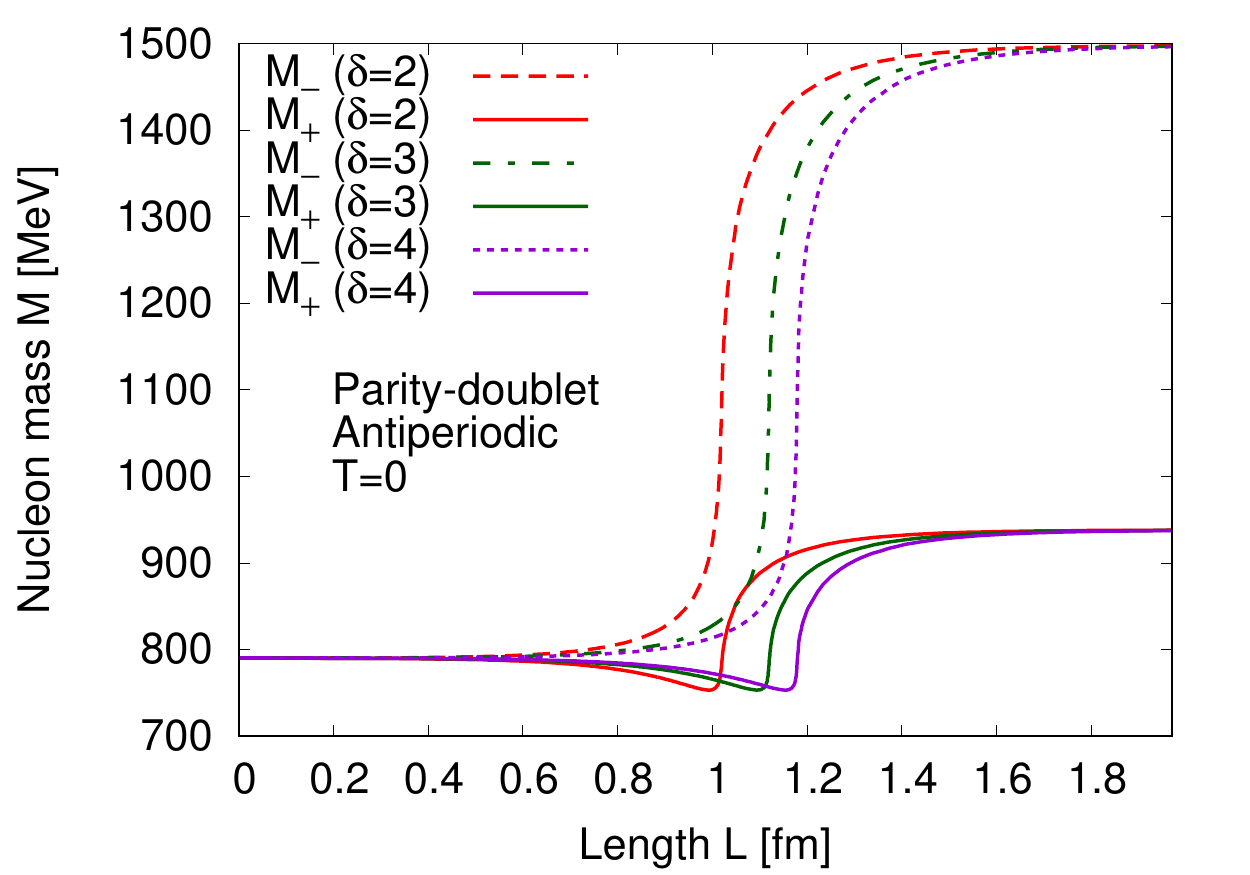}
        \end{center}
    \end{minipage}
    \caption{Finite-volume transition of nucleon masses with $\delta=2,3,4$ and antiperiodic boundary condition at $T=0$.
Upper: Walecka model. Lower: Parity-doublet model.}
    \label{apb_LM_delta234}
\end{figure}

The first term of Eq.~(\ref{Omega_V}) still includes the vacuum energy with the ultraviolet divergence, but by using a regularization scheme, we can estimate a finite energy shift by the finite-volume effect, that is, the Casimir energy.
For the antiperiodic boundary condition, the Casimir energy for massive fermions at zero temperature is given by \cite{Mamaev:1980jn,CougoPinto:1996dr,Elizalde:2002wg,Abreu:2017lxf}
\begin{equation}
\frac{\Omega_\mathrm{Cas}^\mathrm{ap}(L)}{V} = \sum_{i} \gamma_i \sum_{n=1}^\infty (-1)^n \left( \frac{M_i}{n \pi L} \right)^2 K_2(n M_i L),
\label{Casimir_antiperiodic}
\end{equation}
where $K_2$ is the modified Bessel function.
For the periodic boundary condition, 
\begin{equation}
\frac{\Omega_\mathrm{Cas}^\mathrm{p}(L)}{V} = \sum_{i} \gamma_i \sum_{n=1}^\infty \left( \frac{M_i}{n \pi L} \right)^2 K_2(n M_i L).
\label{Casimir_periodic}
\end{equation}
The convergence of this expansion by the modified Bessel function may be practically important.
The function is exponentially damping as $K_2(x) \approx \sqrt{\frac{\pi}{2x}} \mathrm{e}^{-x}$ when $x$ is large enough.
As a result, at a large volume $L$, the Casimir energy is suppressed, as intuitively expected.
Also, the contribution from larger $n$ terms in the summation can be neglect.
In this work, we set the summation up to $n=5$ for numerical calculations.
Then, the error from this truncation is estimated to be at worst $\mathcal{O}$(1\%) due to the factor $1/n^2$.

\section{Numerical results} \label{Sec_3}
\subsection{Finite-$L$ transition with antiperiodic boundary}

The finite-volume effects from the antiperiodic boundary condition are similar to effects from finite temperature.
The leading ($n=1$) term of the Casimir energy in Eq.~(\ref{Casimir_antiperiodic}) has the minus sign for the thermodynamic potential.
For a small $L$, the term is dominated by the second term of $K_2(x) = 2/x^2 - 1/2 + \mathcal{O} (x^2)$ and it is proportional to $M_i^2$.
For this reason, smaller nucleon masses by modification of the $\bar{\sigma}$ mean field are favored, which corresponds to the restoration of chiral symmetry in the parity-doublet model.

\begin{figure}[tb!]
    \begin{minipage}[t]{1.0\columnwidth}
        \begin{center}
            \includegraphics[clip, width=1.0\columnwidth]{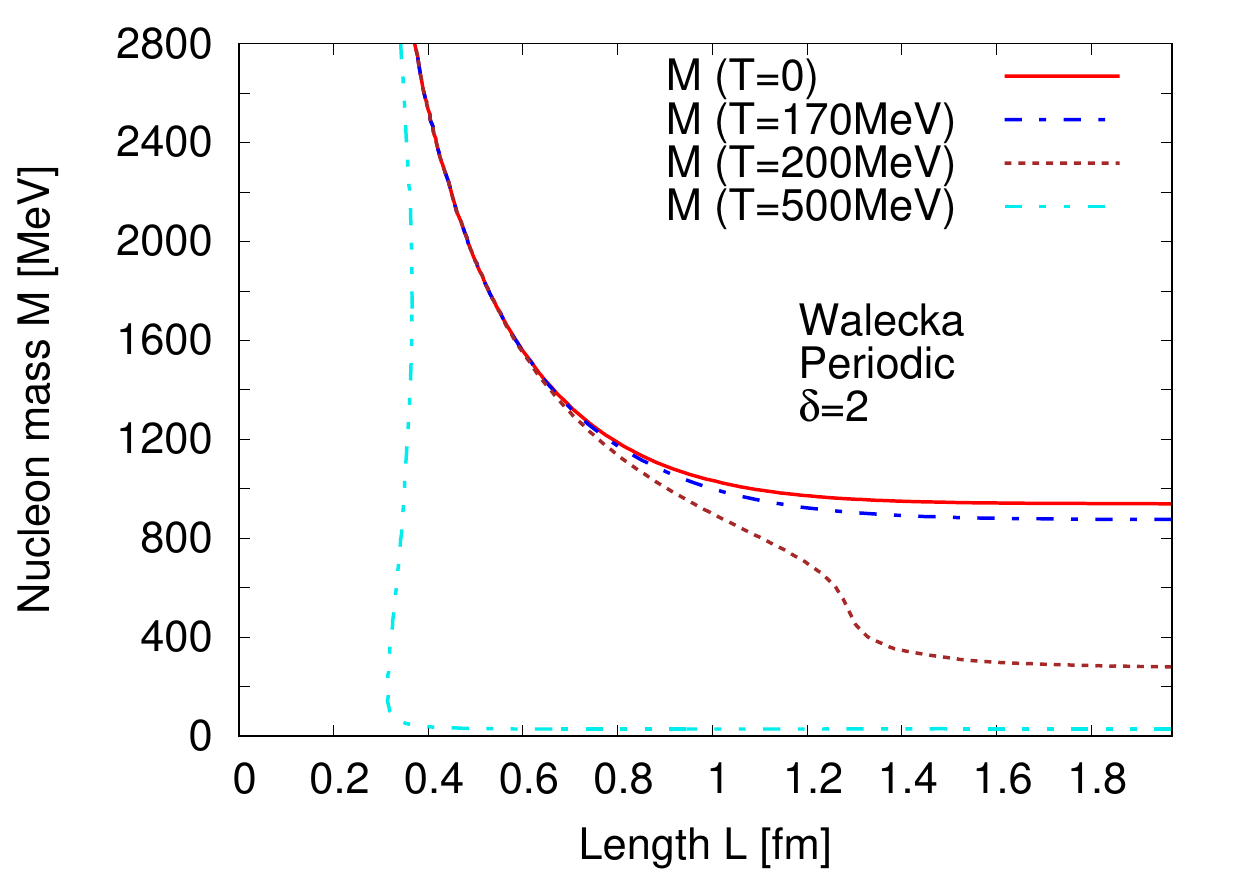}
        \end{center}
    \end{minipage}
    \begin{minipage}[t]{1.0\columnwidth}
        \begin{center}
            \includegraphics[clip, width=1.0\columnwidth]{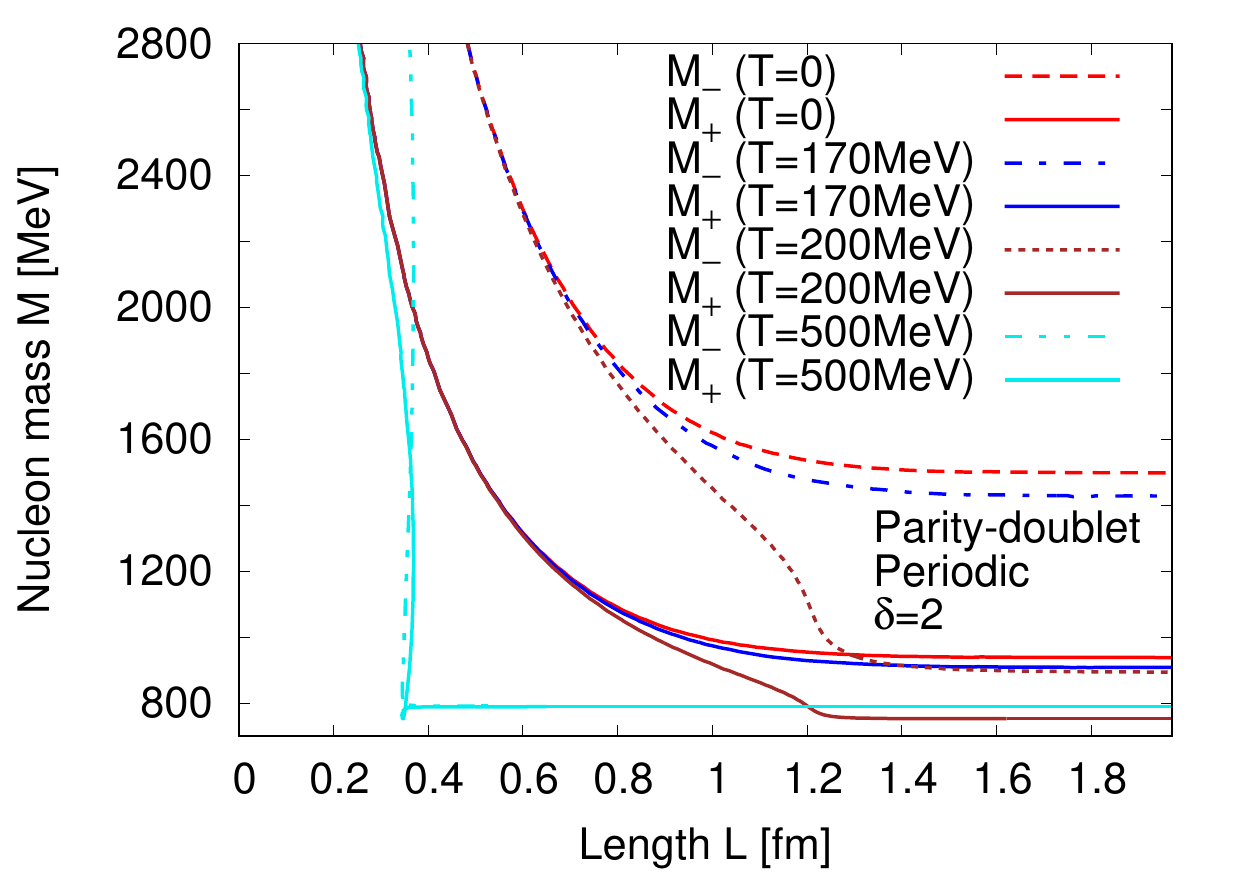}
        \end{center}
    \end{minipage}
    \caption{Finite-volume transition of nucleon masses with $\delta=2$ and periodic boundary condition.
Upper: Walecka model. Lower: Parity-doublet model.}
    \label{pb_LM_delta2}
\end{figure}

\begin{figure}[tb!]
    \begin{minipage}[t]{1.0\columnwidth}
        \begin{center}
            \includegraphics[clip, width=1.0\columnwidth]{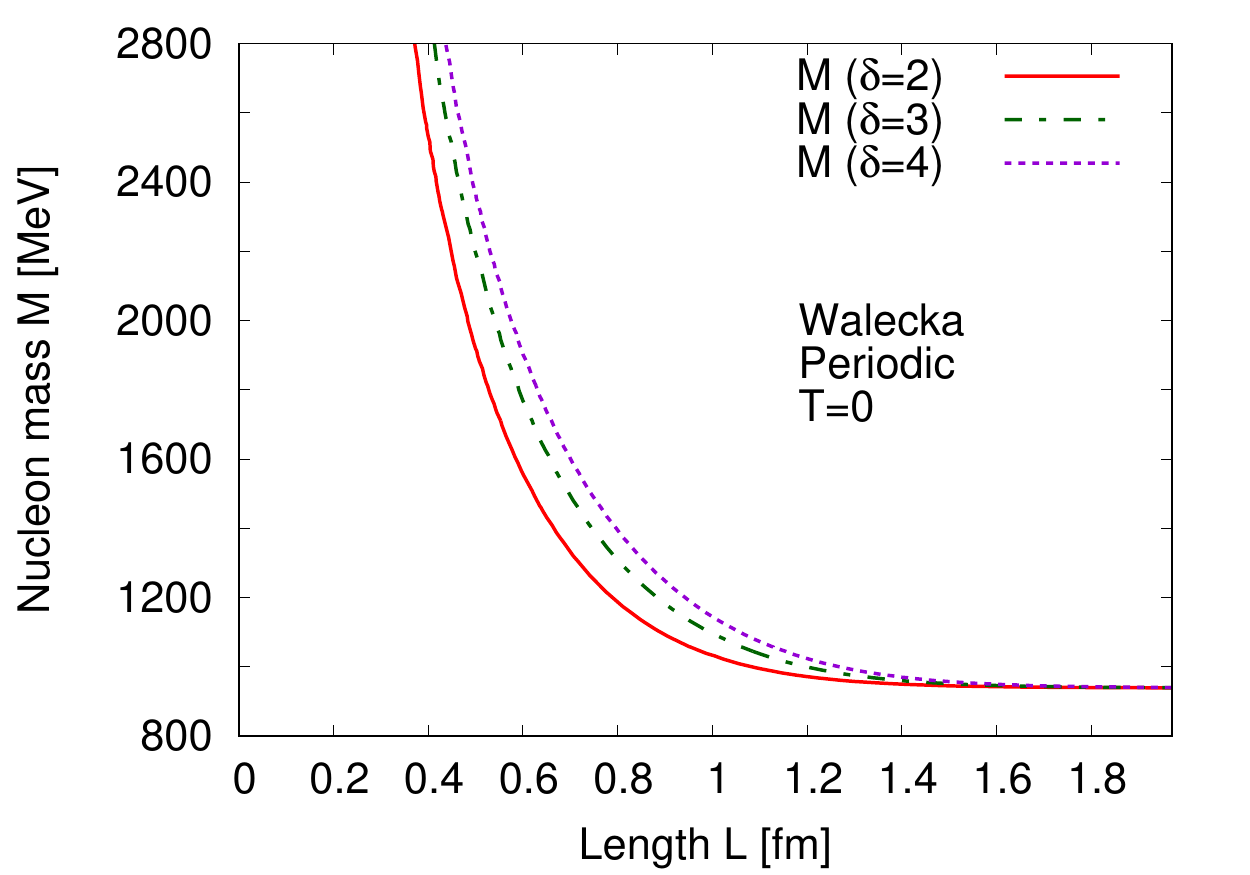}
        \end{center}
    \end{minipage}
    \begin{minipage}[t]{1.0\columnwidth}
        \begin{center}
            \includegraphics[clip, width=1.0\columnwidth]{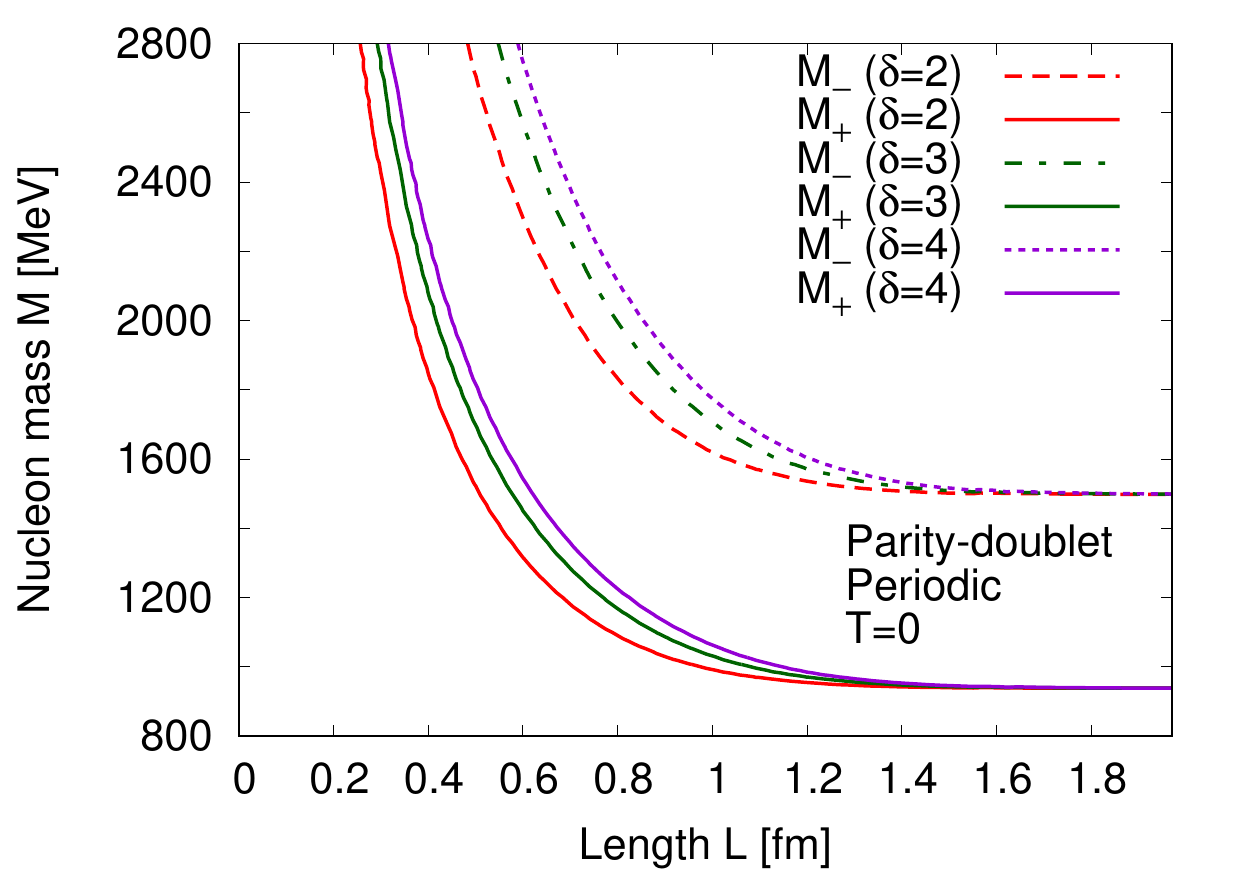}
        \end{center}
    \end{minipage}
    \caption{Finite-volume transition of nucleon masses with $\delta=2,3,4$ and periodic boundary condition at $T=0$.
Upper: Walecka model. Lower: Parity-doublet model.}
    \label{pb_LM_delta234}
\end{figure}

In Figs.~\ref{apb_LM_delta2} and \ref{apb_LM_delta234}, we show the $L$ dependence of nucleon masses in the two models with the antiperiodic boundary condition.
From these figures, our findings are as follows:
\begin{enumerate}
\item In the Walecka model, as $L$ gets smaller, the nucleon mass decreases as shown in the upper panel of Fig.~\ref{apb_LM_delta2}.
This behavior is induced by the increase of $\bar{\sigma}$ by the finite-volume effect.
At the small $L$ limit, the nucleon mass goes to zero.
\item In the parity-doublet model, in the large $L$ region, the masses ($M_+$ and $M_-$) of the nucleon doublet split as shown in the lower panel of Fig.~\ref{apb_LM_delta2}, which is consistent with those in the infinite-volume limit.
As $L$ gets smaller, the masses degenerate, which is induced by the chiral symmetry restoration (or the reduction of $\bar{\sigma}$) by the finite-volume effect.
In the small $L$ region, the nucleon masses agree with the chiral-invariant mass $m_0$.
Around the transition length, the mass splitting in $L \sim 0.8 \ \mathrm{fm}$ is dominated in the linear $\bar{\sigma}$ term since $\bar{\sigma}$ is finite but small.
In the $L$ larger than the transition length, both the masses are lifted up by the $\bar{\sigma}^2$ term with a large $\bar{\sigma}$ value.
\item In any case, at $T$=0, the transition length is about $L \sim 1 \ \mathrm{fm} \sim 0.005 \ \mathrm{MeV^{-1}}$.
This energy scale is comparable to that of the chiral condensate (approximately $ 200 \ \mathrm{MeV}$).
\item With increasing temperature, the transition length is shifted to the larger $L$.
This is because chiral symmetry is partially restored by thermal effects, and the nucleon masses also decrease.
\item In Fig.~\ref{apb_LM_delta234}, we compare the nucleon masses with different compactified dimensions ($\delta=2,3,4$).
In both the models, as $\delta$ increases, the transition length becomes larger.
This is because the finite-volume effect gets stronger by increasing the number of compactified dimensions. 
\end{enumerate}

\subsection{Finite-$L$ transition with periodic boundary}

In contrast to the antiperiodic boundary, the finite-volume effects from the periodic boundary condition lead to a characteristic behavior.
The Casimir energy in Eq.~(\ref{Casimir_periodic}) has the plus sign for the thermodynamics potential, and eventually it is proportional to $-M_i^2$, using $K_2(x) = 2/x^2 - 1/2 + \mathcal{O} (x^2)$ for a small $x$.
For this reason, larger nucleon masses by modification of the $\bar{\sigma}$ mean field are energetically favored.
This corresponds to the increase of the chiral condensate in QCD, which is originally induced by the domination of an infrared quark momentum [or the momentum ``zero mode" as Eq.~(\ref{pz_peri}) for $l=0$].
Such a catalysis of chiral symmetry breaking by the periodic boundary condition for fermions has been observed also from other chiral effective models (e.g., see Refs.~\cite{Song:1990dm,Kim:1994es,Braun:2005gy,Braun:2005fj,Palhares:2009tf,Tripolt:2013zfa}).\footnote{As an analogous phenomenon, the chiral symmetry breaking induced by magnetic fields, which is the so-called magnetic catalysis, is well known \cite{Klevansky:1989vi,Suganuma:1990nn,Gusynin:1994re,Gusynin:1995nb,Shovkovy:2012zn,Miransky:2015ava}.}

\begin{figure}[t!]
    \begin{minipage}[t]{1.0\columnwidth}
        \begin{center}
            \includegraphics[clip, width=1.0\columnwidth]{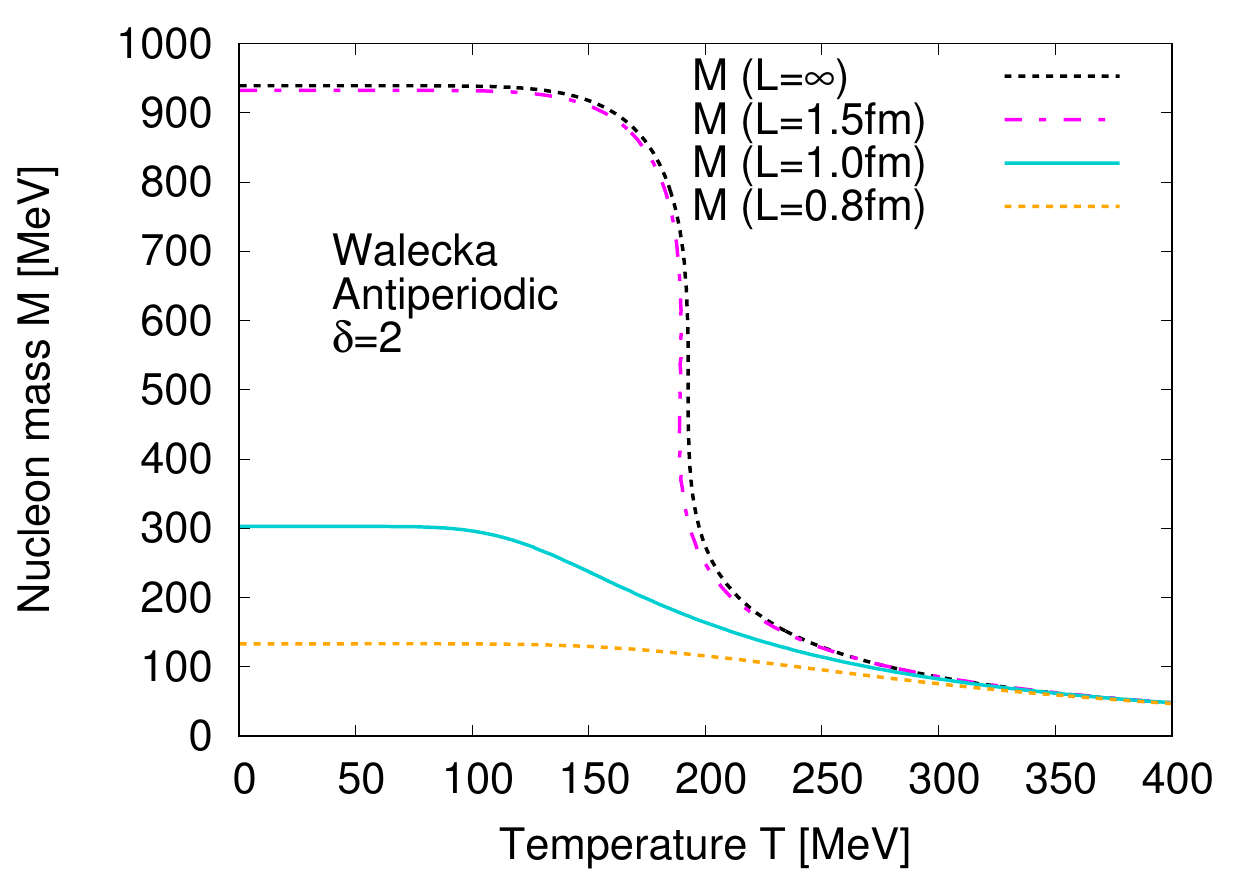}
        \end{center}
    \end{minipage}
    \begin{minipage}[t]{1.0\columnwidth}
        \begin{center}
            \includegraphics[clip, width=1.0\columnwidth]{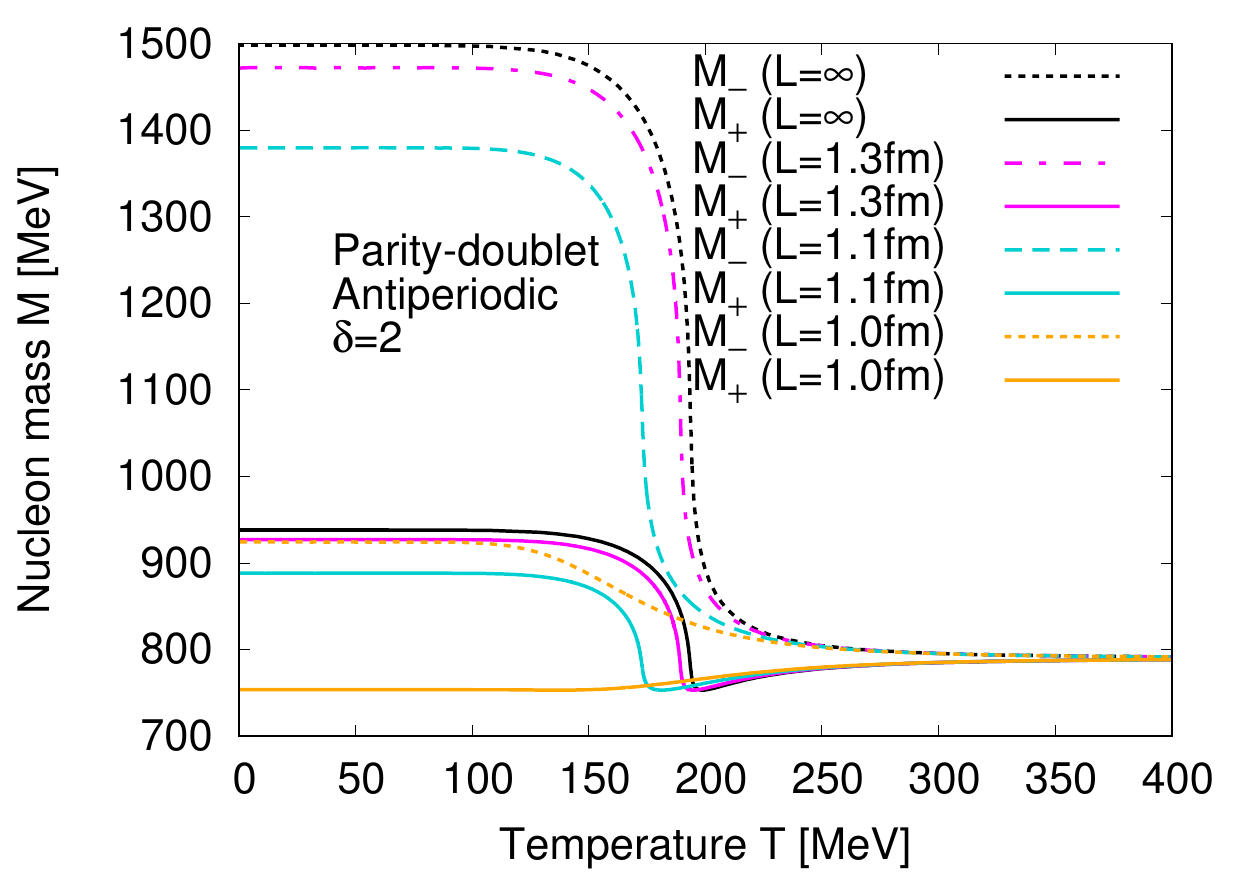}
        \end{center}
    \end{minipage}
    \caption{Finite-temperature transition of nucleon masses with $\delta=2$ and antiperiodic boundary condition. Upper: Walecka model. Lower: Parity-doublet model.}
    \label{apb_TM_delta2}
\end{figure}

\begin{figure}[t!]
    \begin{minipage}[t]{1.0\columnwidth}
        \begin{center}
            \includegraphics[clip, width=1.0\columnwidth]{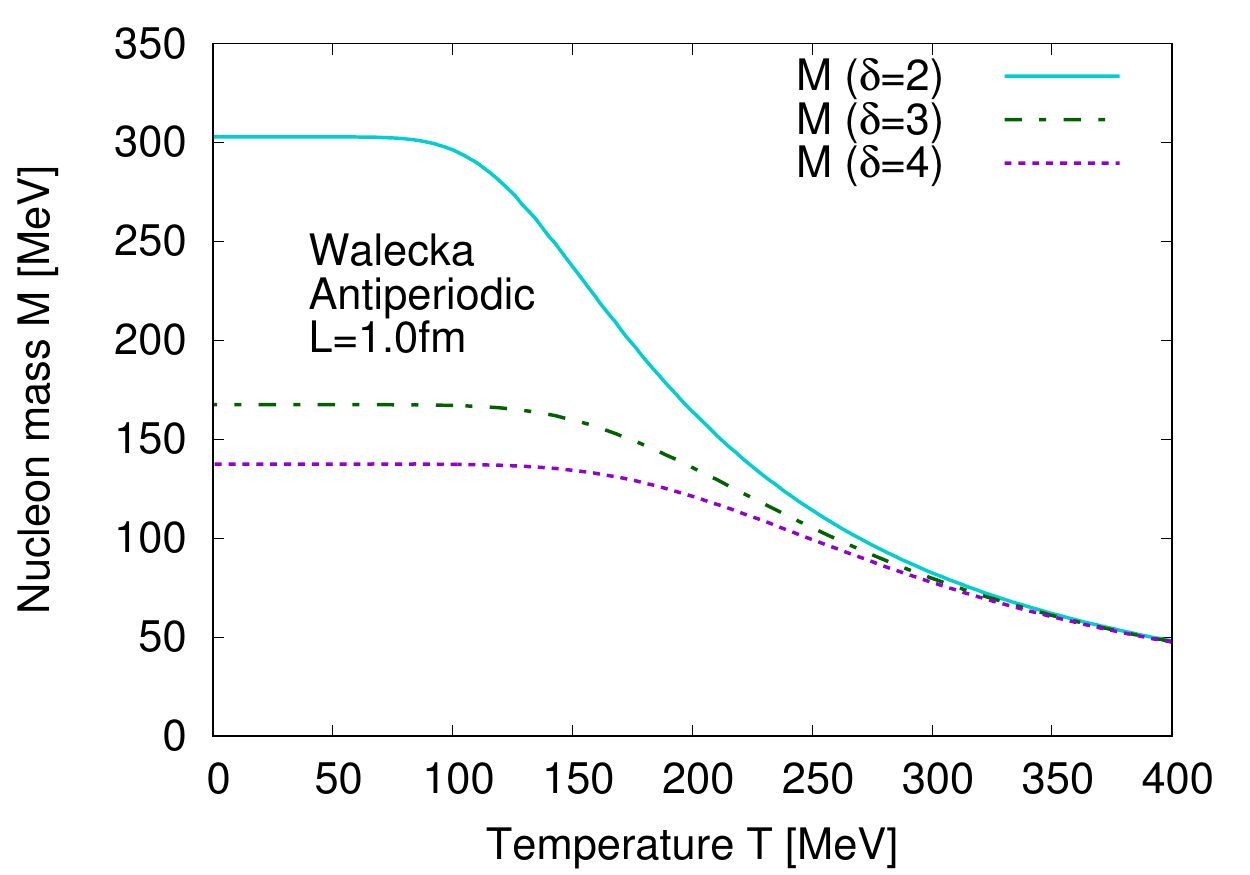}
        \end{center}
    \end{minipage}
    \begin{minipage}[t]{1.0\columnwidth}
        \begin{center}
            \includegraphics[clip, width=1.0\columnwidth]{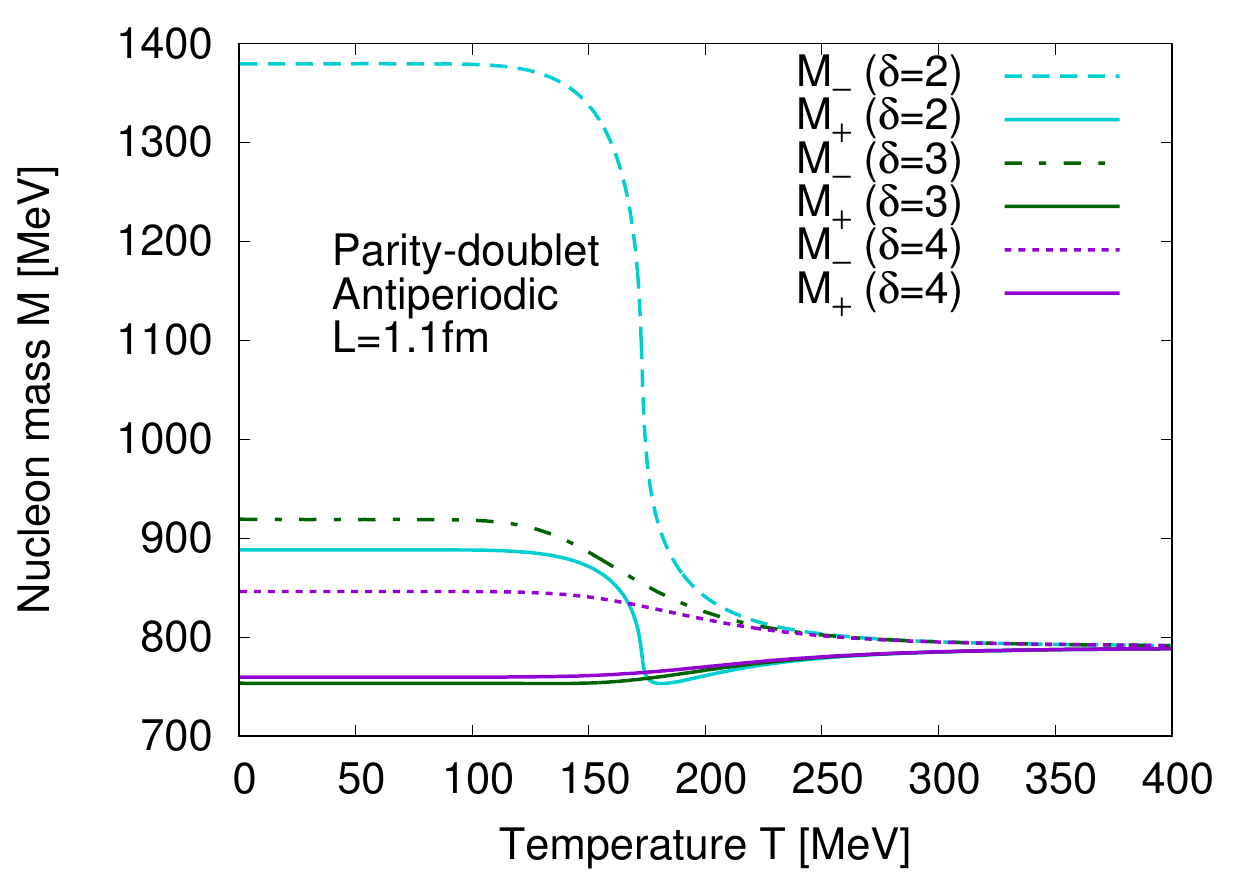}
        \end{center}
    \end{minipage}
    \caption{Finite-temperature transition of nucleon masses with $\delta=2,3,4$ and antiperiodic boundary condition. Upper: Walecka model at $L=1.0 \ \mathrm{fm}$. Lower: Parity-doublet model at $L=1.1 \ \mathrm{fm}$.}
    \label{apb_TM_delta234}
\end{figure}

In Figs.~\ref{pb_LM_delta2} and \ref{pb_LM_delta234}, we show the $L$ dependence of nucleon masses with the periodic boundary condition at a fixed $T$.
From these figures, our findings are as follows:
\begin{enumerate}
\item In the Walecka model, as $L$ gets smaller, the nucleon mass increases as shown in the upper panel of Fig.~\ref{pb_LM_delta2}.
This is corresponding to the enhancement of the chiral symmetry breaking (or the decrease of $\bar{\sigma}$) induced by finite-volume effects with the periodic boundary condition.
\item In the parity-doublet model, as $L$ gets smaller, the masses of both the nucleons increase as shown in the lower panel of Fig.~\ref{pb_LM_delta2}.
This is also corresponding to the enhancement of chiral symmetry breaking (or the increase of $\bar{\sigma}$).
\item For both the models, in the large-$L$ region, with increasing temperature, the nucleon masses decrease by the chiral symmetry restoration.
On the other hand, in the small-$L$ region below $L \sim 0.6 \mathrm{fm} \sim 0.003 \ \mathrm{MeV}^{-1}$, the nucleon masses are independent of temperature.
This is because the nucleon mass shifts are dominated by the finite-volume effects with a large scale (approximately $330 \ \mathrm{MeV}$) and thermal effects relatively do not contribute to the nucleons.  
\item In Fig.~\ref{pb_LM_delta234}, we compare the different compactified dimensions ($\delta=2,3,4$).
In both models, as $\delta$ increases, the nucleon masses also increase.
This is because the chiral symmetry breaking is enhanced by increasing the number of compactified dimensions. 
\end{enumerate}

\subsection{Finite-$T$ transition with antiperiodic boundary}

\begin{figure}[t!]
    \begin{minipage}[t]{1.0\columnwidth}
        \begin{center}
            \includegraphics[clip, width=1.0\columnwidth]{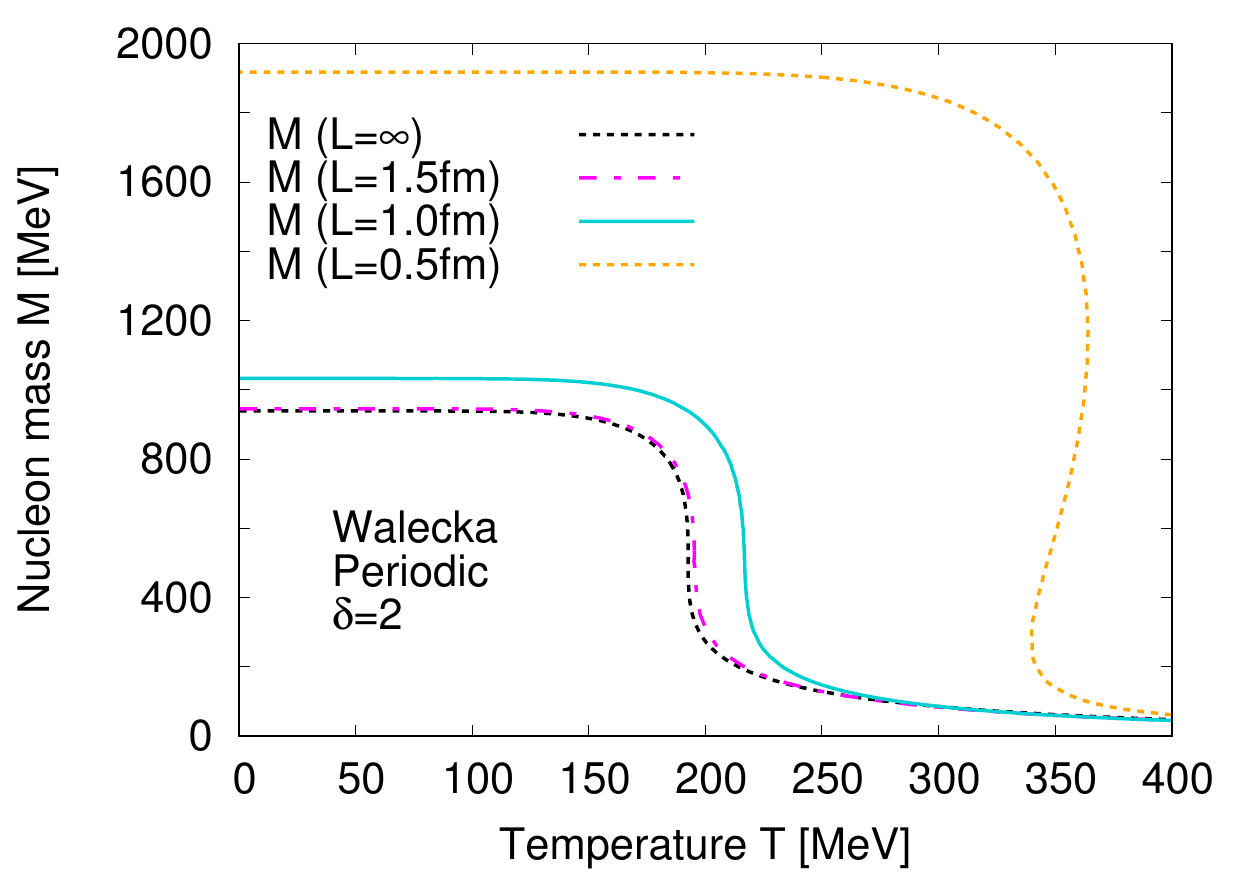}
        \end{center}
    \end{minipage}
    \begin{minipage}[t]{1.0\columnwidth}
        \begin{center}
            \includegraphics[clip, width=1.0\columnwidth]{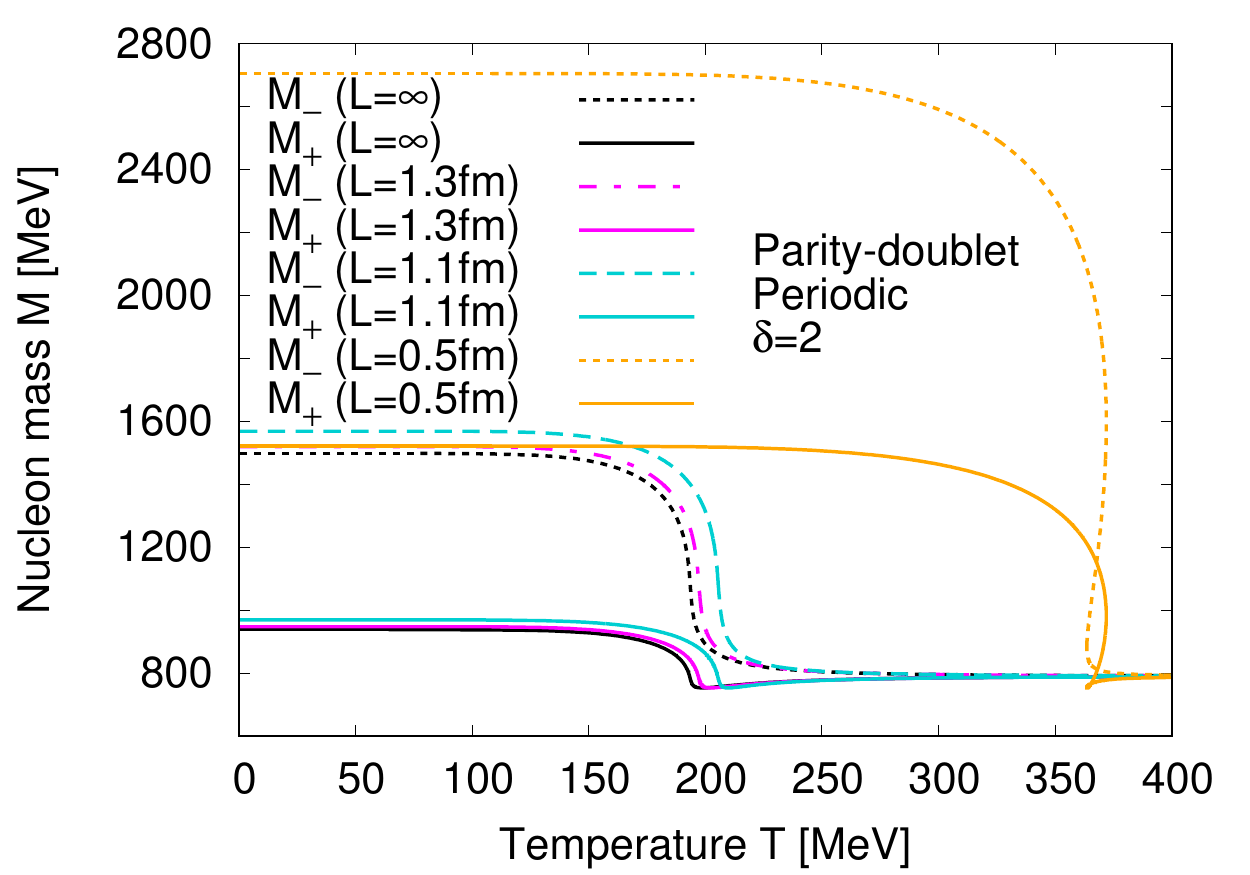}
        \end{center}
    \end{minipage}
    \caption{Finite-temperature transition of nucleon masses with $\delta=2$ and periodic boundary condition. Upper: Walecka model. Lower: Parity-doublet model.}
    \label{pb_TM_delta2}
\end{figure}

\begin{figure}[t!]
    \begin{minipage}[t]{1.0\columnwidth}
        \begin{center}
            \includegraphics[clip, width=1.0\columnwidth]{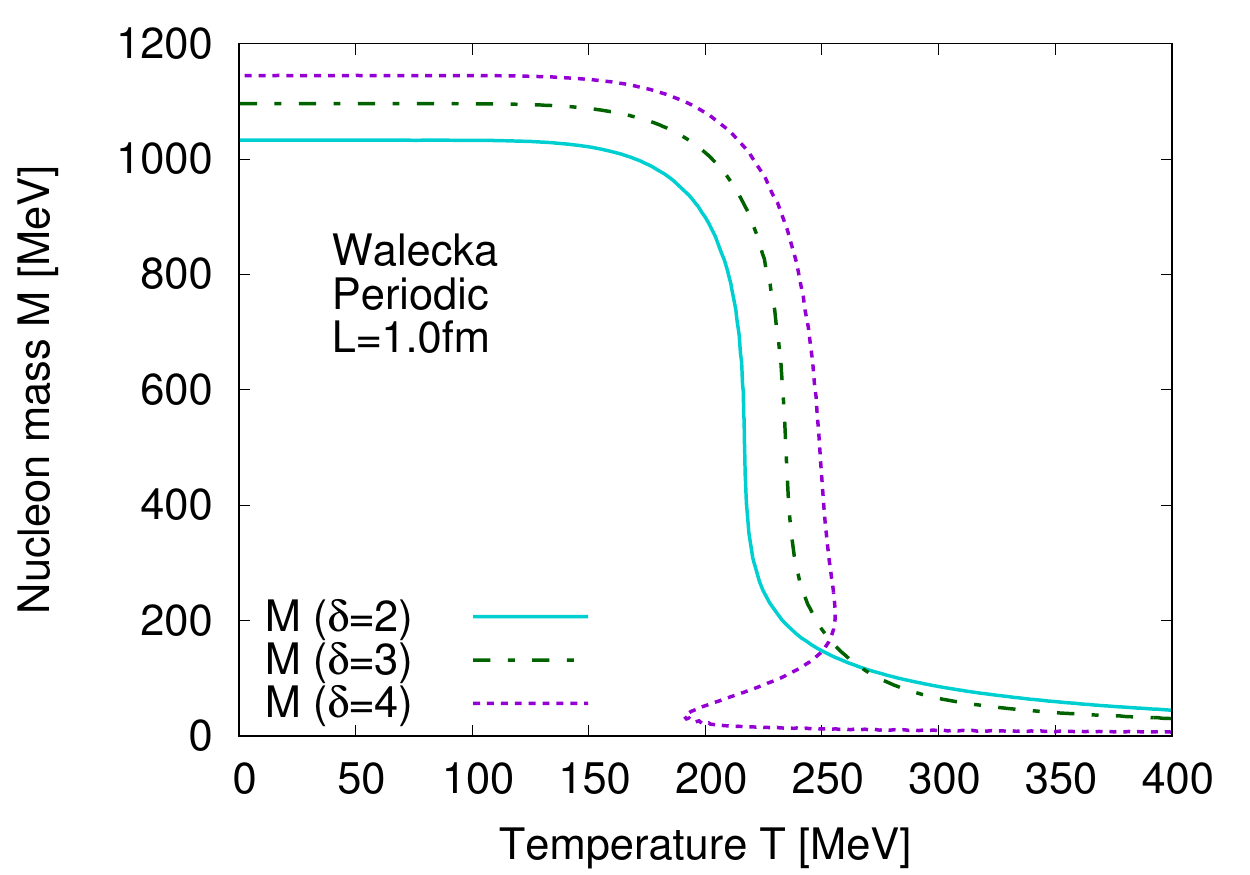}
        \end{center}
    \end{minipage}
    \begin{minipage}[t]{1.0\columnwidth}
        \begin{center}
            \includegraphics[clip, width=1.0\columnwidth]{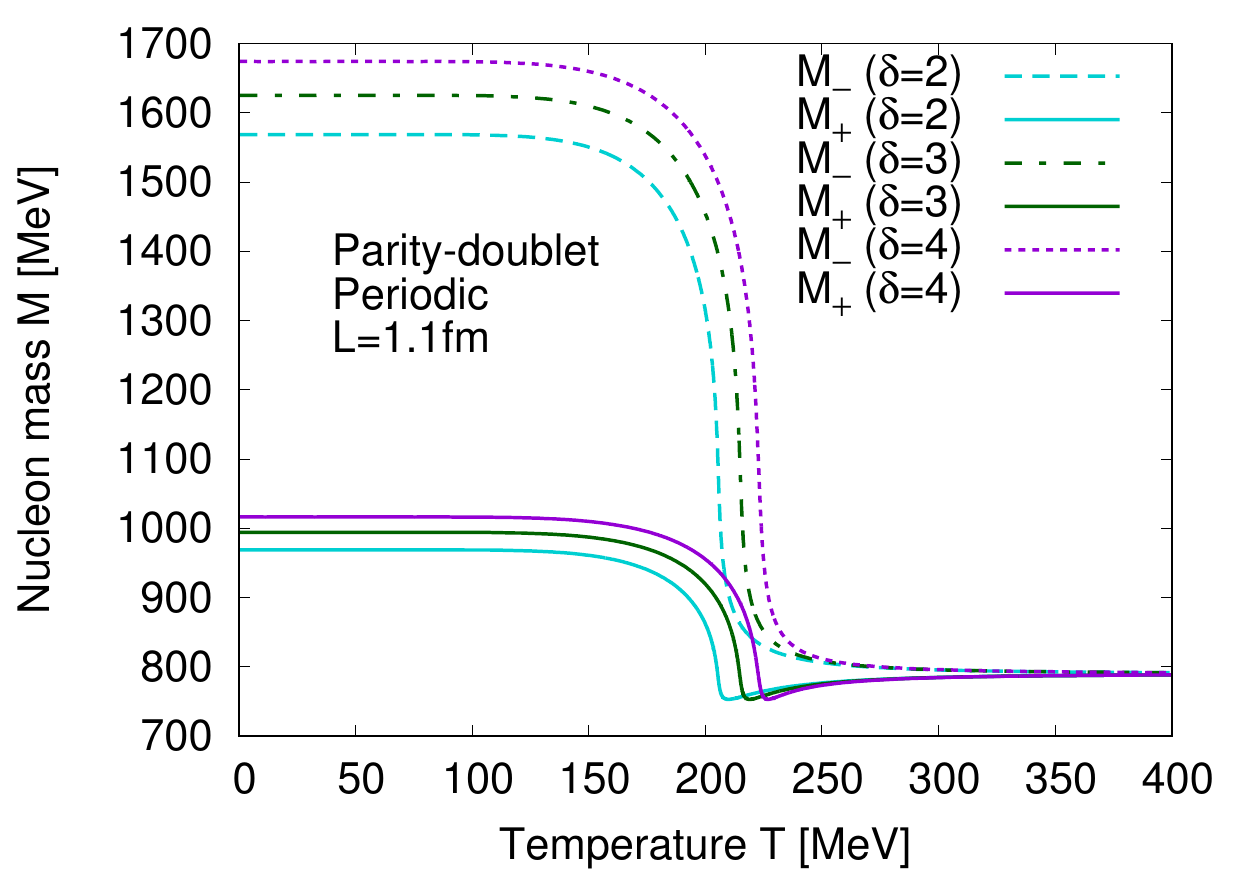}
        \end{center}
    \end{minipage}
    \caption{Finite-temperature transition of nucleon masses with $\delta=2,3,4$ and periodic boundary condition. Upper: Walecka model at  $L=1.0 \ \mathrm{fm}$. Lower: Parity-doublet model at $L=1.1 \ \mathrm{fm}$.}
    \label{pb_TM_delta234}
\end{figure}

In this and the next sections, we investigate finite-volume effects for thermal phase transitions.
Notice that, for our parameters, the thermal transition for $\bar{\sigma}$ of the Walecka model at infinite volume is a crossover.\footnote{If the coupling constant $g_\sigma$ is larger, the thermal phase transition can be first order.
In this case, finite-volume effects have already been studied in Ref.~\cite{Abreu:2017lxf}.
Therefore, in this work, we focus on the crossover transition.}
Furthermore, the order for the parity-doublet model is also a crossover.\footnote{If we use another setup for the parity-doublet model, the thermal phase transition could be first order.
For such a situation, see Appendix~\ref{App_2}, in which a six-point scalar vertex is introduced.}

In Figs.~\ref{apb_TM_delta2} and \ref{apb_TM_delta234}, we show the temperature dependence of the nucleon masses at a fixed $L$ with the antiperiodic boundary condition.
As we mentioned, the finite-volume effect from antiperiodic boundary condition is similar to the finite temperature effect.
From these figures, our findings are as follows:
\begin{enumerate}
\item In the Walecka model, as the $L$ decreases, the nucleon mass at low temperature (in the chiral-broken phase) also decreases.
The order of the phase transition is still a crossover.
\item In the parity-doublet model, as $L$ gets smaller, the nucleon masses at low temperature decrease, and the transition temperature also decreases.
\item In Fig.~\ref{apb_TM_delta234}, we compare the nucleon masses with different compactified dimensions ($\delta=2,3,4$).
In both the models, as $\delta$ increases, the nucleon masses decrease, and the transition temperature also decreases.
\end{enumerate}

\subsection{Finite-$T$ transition with periodic boundary}
In Figs.~\ref{pb_TM_delta2} and \ref{pb_TM_delta234}, we show the temperature dependence of nucleon masses at a fixed $L$ with the periodic boundary condition.
From these figures, our findings are as follows:
\begin{enumerate}
\item For both the models, as the $L$ decreases, the nucleon mass in the low-temperature phase increases as the result of chiral symmetry breaking, and the transition temperature also increases.
The order of the phase transition becomes first order in small volume $L$.

\item In Fig.~\ref{pb_TM_delta234}, we compare the nucleon masses with different compactified dimensions ($\delta=2,3,4$).
In both the models, as $\delta$ increases, the nucleon masses increase, and the transition temperature also increases.
The order of the phase transition becomes first order in the more compactified case.
Thus, a larger $\delta$ provides more substantial finite-volume effects.
\end{enumerate}

\section{Conclusion and outlook}  \label{Sec_4}
In this work, we have shown finite-volume effects for the nucleon masses in terms of the chiral partner structure.
This is the first attempt to apply the nucleon parity-doublet model with the mirror assignment~\cite{Detar:1988kn} to finite-volume systems,
in which we introduced the finite-volume effects as the Casimir effects.
For the antiperiodic boundary, the finite-volume effects are similar to the effects from finite temperature, and chiral symmetry is restored in smaller volume $L$.
For the periodic boundary, the finite-volume effect lifts the masses of nucleons for small volume $L$.
In addition, the thermal crossover transition at large $L$ could change to the first order in the small-$L$ region.

The advantage of the parity-doublet model is to separate the origins of the nucleon masses into the two components: the chiral condensate (or $\bar{\sigma}$ mean field) and the chiral-invariant mass $m_0$.
While the chiral condensate is well known in QCD, the origin of the chiral-invariant mass and its precise value are still unknown.
As studied in Refs.~\cite{Aarts:2015mma,Aarts:2017rrl,Aarts:2018glk}, investigating the degeneracy between the chiral partners at finite temperature on the lattice is one of the powerful approaches to study $m_0$. 
Here, we newly suggest that small-volume systems will be useful for studying the origin of nucleon masses.
We emphasize that this situation is similar to finite temperature/density but essentially different:
while at finite temperature/density we always need to take into account the excitation of hadronic/nuclear matter,
in the finite volume we could interpret the behavior of the chiral symmetry breaking in a different manner and would extract the value of the chiral invariant mass more clearly.
Our novel setup will provide a new motivation in both the lattice QCD simulations and model studies.

As mentioned in the Introduction, the finite-volume effect as an artifact in lattice QCD simulations is well recognized, but ``the origin" of finite-volume effects for nucleon masses is still not elucidated.
One may try to interpret it using the chiral perturbation theory with baryons in finite volume.
Our scenario in terms of the parity-doublet model and mean-field approximation is an alternative/additional interpretation for nucleon mass shifts measured in small volume.
Furthermore, to understand finite-volume effects for negative-parity excited states such as $N^\ast(1535)$ might be more a challenging task.
This work is the first suggestion of finite-volume effects in terms of the chiral partner structure.

We comment on the dependence on the number of compactified dimensions (denoted as $\delta$).
Here, while the $\delta=4$ is usual setup in lattice QCD simulations, the setup in $\delta=2$ or $3$ is unusual but might be interesting.
For example, $\delta =2$ is related to the Casimir effect in QCD vacuum with two parallel plates.
Under such environments, to study various physical quantities such as the chiral condensate and hadron masses on both the continuum and lattice theory is also interesting.
This is because we can tune $\delta$ as a new parameter and examine the dependence of observables, which is a different advantage from studies at finite temperature.
In particular, we emphasize that our numerical results show a relevant $\delta$ dependence of the nucleon mass shifts.
Furthermore, the temperature at which the degeneracy between the partners occurs, the ``degeneracy temperature" (which might be related to the pseudocritical temperature in QCD), and the ``degeneracy length" are relevantly modified.

Usually, the Walecka and parity-doublet models are useful for studying finite-density systems, namely nuclear matter.
Investigation of the finite-volume (and Casimir) effects for the nuclear matter is left for future works \cite{Ishikawa:prep}.
In this situation, we can consider not only the $\omega$ mean field but also other mean fields: the competition between the ``usual" nuclear matter with the homogeneous $\sigma$ and $\omega$ mean fields and the ``anomalous" phase with the inhomogeneous chiral condensate (or the so-called chiral density waves) is also interesting, as discussed in Refs.~\cite{Heinz:2013hza,Takeda:2018ldi}.

In the framework of the parity-doublet model, other additional degrees of freedom can be included.
For example, the parity-doublet model taking into account the $\Delta$ isobar (the so-called {\it chiral quartet scheme}) was first suggested in Ref.~ \cite{Jido:1999hd}, and the properties of symmetric and asymmetric nuclear matter including $\Delta$ isobars were investigated in Ref.~\cite{Takeda:2017mrm}.
Its thermal behaviors were investigated in Ref.~\cite{Sasaki:2017glk}.
Moreover, the extension of the parity-doublet model to flavor SU(3) would also be interesting \cite{Nemoto:1998um,Chen:2008qv,Chen:2009sf,Chen:2010ba,Chen:2011rh,Steinheimer:2011ea,Nishihara:2015fka,Olbrich:2015gln,Dmitrasinovic:2016hup,Olbrich:2017fsd,Sasaki:2017glk}.
To extend the parity doublet structure to the other symmetries \cite{Catillo:2018cyv} would also be useful for understanding the relation between baryon properties and chiral symmetry.

\section*{Acknowledgments}
The authors are grateful to Daiki Suenaga for giving us helpful comments on the parity-doublet model.
This work was supported by JSPS KAKENHI (Grant No. JP17K14277).
K. N. is supported partly by the Grant-in-Aid for Japan Society for the Promotion of Science Research Fellow (No. 18J11457).
K. S. is supported by MEXT as ``Priority Issue on Post-K computer" (Elucidation of the Fundamental Laws and Evolution of the Universe) and Joint Institute for Computational Fundamental Science.

\appendix
\section{Derivation of Casimir effects} \label{App_1}
In this Appendix, we introduce the regularization scheme, which is essential for the definition of the Casimir effect.
Here, we summarize the regularization by the zeta function.

To define the Casimir energy from the thermodynamic potential, we use the Epstein-Hurwitz inhomogeneous zeta function $Y(s)$ (see Ref.~\cite{Elizalde:2012zza} for a textbook).
For generality, we consider $D$-dimensional space-time with compactified $\delta$-dimensional space-time.
For example, the theory on the $3+1$ dimensional space-time at finite temperature corresponds to $D=4$ and $\delta = 1$.

We consider the potential from the partition function for fields with a mass $M$ and chemical potential $\mu^\ast$
\be{
\frac{\Omega}{V}
=
- \frac{\gamma}{\beta L_1 \cdots L_{\delta-1}}
\sum_{l_0, \cdots, l_{\delta-1} = -\infty} ^{\infty}
\int_{-\infty} ^\infty \frac{dq^{D-\delta}}{(2\pi)^{D-\delta}}
f (q,p),
}
where $V$, $\beta=1/T$, $\gamma$, and $L_i$ are the volume, inverse temperature, degeneracy factor, and size of compactified dimensions, respectively.
$q$ and $p$ are the continuous and discretized momenta, respectively.
$l_0, \cdots, l_{\delta-1}$ are the mode indices for the discretized momenta, and
\be{
f (q,p)
=
\ln \left[ \sum_{i=1} ^{\delta-1}(p_{i})^2 + \sum_{j=1} ^{D-\delta}(q_j)^2 + {(p_0 - i\mu^*)}^2 +M^2 \right].
}
Here, $p_{i} = \frac{2\pi}{L_i} \left( l_i + \frac{\alpha_i}{2} \right)$, $q_j$, and $p_0 = \frac{2\pi}{\beta} \left( l_0 + \frac{\alpha_0}{2} \right)$ are the discretized momentum in the $i$ th dimension, continuous momentum in the $j$ th dimension, and time component, respectively.
We also introduce the parameter $\alpha_i$ for denoting a boundary condition.
For the antiperiodic boundary condition, this symbol takes $\alpha_i = 1$, and the periodic boundary condition corresponds to $\alpha_i=0$.
This potential diverges by the sum of the contribution from high-momentum modes.
We take analytical continuation for $D-\delta$ using a regularization with the zeta function.
After the regularization, we perform the $D-\delta$-dimensional integration in the polar coordinates using the relation $\int_0 ^\infty t^r(1+t)^s\mathrm{d}t = \frac{\Gamma(1+r)\Gamma(-1-r-s)}{\Gamma(-s)}$.

Then the potential $\Omega$ is represented by
\be{
\Omega(\beta,\{L_i\},\mu^\ast)
=
\frac{\gamma V}{\beta L_1\times \cdots \times L_{\delta-1}}
\left[\pdif{Y}{s}\right]_{s=0}, \label{eq_A3}
}
where the function $Y(s)$ is introduced as
\beqnn{
Y(s)
&=&
\frac{1}{(4\pi)^{(D-\delta)/2}}
\frac{\Gamma(\nu)}{\Gamma(s)}
\nonumber\\
&&
\times\sum_{l_0, \cdots, l_{\delta-1}=-\infty} ^{\infty}
\left[
\left(
\frac{2\pi}{\beta}
\left(
l_0 + \frac{\alpha_0}{2}
\right)
-i\mu^*
\right)^2\right.
\nonumber\\
&&
+\left.\sum_{i=1} ^{\delta-1}
\left(
\frac{2\pi}{L_i}
\left(
l_i + \frac{\alpha_i}{2}
\right)
\right)^2
+M ^2
\right]^{-\nu}, \label{eq_A4}
}
with the parameter $\nu \equiv s-\frac{D-\delta}{2}$.


We expand the function $Y(s)$ by the modified Bessel function $K_n(x)$ for the regularization.
The expansion is represented with the parameters $a_i$ and $b_i$ as follows \cite{Kirsten:1994yp}:
\beqnn{
&&\sum_{l_0, \cdots, l_{\delta-1} = -\infty} ^{\infty}
\left[
\sum_{i = 0} ^{\delta-1}
a_i\left(l_i - b_i\right)^2 +M^2
\right] ^{-\nu}\nonumber\\
&=&
\frac{2}{\Gamma(\nu)}\frac{\pi^{\delta/2}}{\sqrt{a_0 \cdots a_{\delta-1}}}
\left[
\frac{1}{2} \Gamma(s-\frac{D}{2})M^{D-2s}\right.\nonumber\\
&+&
2
\sum_{i=0} ^{\delta-1} 
\sum_{n_i = 1} ^\infty
\mathrm{cos}{\left(2\pi n_ib_i\right)}
\left(
\frac{\pi n_i}{\sqrt{a_i}M}
\right)^{s-\frac{D}{2}}
K_{s-\frac{D}{2}}\left(
\frac{2\pi n_i M}{\sqrt{a_i}}
\right)\nonumber\\
&+&
2^2
\sum_{i < j = 0} ^{\delta-1} 
\sum_{n_i,n_j = 1} ^{\infty} 
\mathrm{cos}(2\pi n_ib_i)
\mathrm{cos}(2\pi n_jb_j)\nonumber\\
&\times&
\left(
\frac{\pi}{M}
\sqrt{
\frac{n_i ^2}{a_i}
+
\frac{n_j ^2}{a_j}
}
\right)^{s-\frac{D}{2}}
K_{s-\frac{D}{2}}\left(
2\pi M
\sqrt{
\frac{n_i ^2}{a_i}
+
\frac{n_j ^2}{a_j}
}
\right)\nonumber\\
&+& \cdots \nonumber\\
&+&
{ 2 }^\delta 
\sum_{n_0, \cdots, n_{\delta-1} = 1} ^\infty
\prod_{i=0} ^{\delta-1}[\mathrm{cos}(2\pi n_i b_i)]\nonumber\\
&\times& \left.
\left(
\frac{\pi}{M }
\sqrt{\sum_{i = 0} ^{\delta-1} \frac{n_i ^2}{a_i}}
\right)^{s-\frac{D}{2}}
K_{s-\frac{D}{2}}\left(
2\pi M
\sqrt{\sum_{i = 0} ^{\delta-1} \frac{n_i ^2}{a_i}}
\right) \right].
}

For our case, $a_i=\left(\frac{2\pi}{L_i}\right)^2$ with $L_0 = \beta$.
The constant $b_i = -\frac{\alpha_i}{2} + i\frac{\mu_i}{\sqrt{a_i}}$ is defined by the boundary condition and the chemical potential.
We introduce the chemical potential with index $\mu_{i\neq 0} = 0$, $\mu_{i= 0} = \mu^*$ for simplicity.
For spatial direction ($i\neq 0$), $b_i$ depends on the boundary condition, $b_{i\neq 0} = -\frac{\alpha_i}{2}$, and for time direction, it also depends on the chemical potential $b_{i= 0} = -\frac{\alpha_0}{2} + i\frac{\mu^*}{\sqrt{a_0}}$.

By using this expansion, we can get the $Y(s)$ explicitly,
\beqnn{ 
&&\frac{\gamma V}{\beta L_1 \times  \cdots \times L_{\delta-1}}Y(s)\nonumber\\
&=&
\frac{\gamma V}{(4\pi)^{\frac{D}{2}}}
\frac{2}{\Gamma(s)}
\left[
\frac{1}{2} \Gamma(s-\frac{D}{2})M ^{D-2s}\right.\nonumber\\
&+&
2
\sum_{i=0} ^{\delta-1} 
\sum_{n_i = 1} ^\infty
(-1)^{n_i\alpha_i}
\mathrm{cosh}(n_iL_{i}\mu_i)
\left(
\frac{n_iL_{i}}{2M}
\right)^{s-\frac{D}{2}}
K_{s-\frac{D}{2}}\left(
{n_i ML_{i}}
\right)\nonumber\\
&+&
2^2
\sum_{i < j = 0} ^{\delta-1} 
\sum_{n_i,n_j = 1} ^{\infty} 
(-1)^{n_i\alpha_i + n_j\alpha_j}
\mathrm{cosh}(n_iL_{i}\mu_i)
\mathrm{cosh}(n_jL_{j}\mu_j)\nonumber\\
&\times&
\left(
\frac{1}{2M}
\sqrt{
n_i ^2 L_{i} ^2
+
n_j ^2 L_{j} ^2
}
\right)^{s-\frac{D}{2}}
K_{s-\frac{D}{2}}\left(
M
\sqrt{
n_i ^2 L_{i} ^2
+
n_j ^2 L_{j} ^2
}
\right)\nonumber\\
&+& \cdots \nonumber\\
&+&
2^\delta 
\sum_{n_0, \cdots, n_{\delta-1} = 1} ^\infty
\prod_{i=0} ^{\delta -1}
(-1)^{n_i\alpha_i}
[\mathrm{cosh}(n_i L_{i}\mu_i)]\nonumber}
\vspace{-13pt}
\beqnn{
&\times& \left.
\left(
\frac{1}{2M}
\sqrt{\sum_{i = 0} ^{\delta-1} n_i ^2 L_{i} ^2}
\right)^{s-\frac{D}{2}}
K_{s-\frac{D}{2}}\left(
M
\sqrt{\sum_{i = 0} ^{\delta-1} n_i ^2 L_{i} ^2}
\right) \right]. \hspace{8mm} \label{eq_A6}
}

To obtain $\left[\pdif{Y}{s}\right]_{s=0}$ of Eq.~(\ref{eq_A3}), we use the relation for any regular function $G(s)$,
\be{
\lim_{s\to 0}
\frac{d}{ds}\frac{G(s)}{\Gamma(s)}=G(0).
}
We also use the property of the Bessel function, $K_{-n}(x) = K_n(x)$, and then the thermodynamic potential (\ref{eq_A3}) can be represented by
\beqnn{
\frac{\Omega}{V}
&=&
\Omega_{0}\nonumber\\
&+&
\frac{2\gamma}{(2\pi)^{\frac{D}{2}}}\left[
2
\sum_{i=0} ^{\delta-1}
\sum_{n_i = 1} ^\infty
(-1)^{n_i\alpha_i}
\mathrm{cosh}(n_iL_{i}\mu_i)\right.\nonumber\\
&\times&
\left(\frac{M}{n_iL_{i}}\right)^{\frac{D}{2}}
K_{\frac{D}{2}}(n_i M L_{i})\nonumber\\
&+& \cdots \nonumber\\
&+&
2^\delta 
\sum_{n_0, \cdots, n_{\delta-1} = 1} ^\infty
\prod_{i=0} ^{\delta-1}
(-1)^{n_i\alpha_i}
[\mathrm{cosh}(n_i L_{i}\mu_i)]\nonumber\\
& \times& \left.
\left(
\frac{M}{\sqrt{\sum_{i = 0} ^{\delta-1} n_i ^2 L_{i} ^2}}
\right)^{\frac{D}{2}}
K_{\frac{D}{2}}\left(
M
\sqrt{\sum_{i = 0} ^{\delta-1} n_i ^2 L_{i} ^2}
\right) \right]. \nonumber\\
\label{Cas_ene_gene}
}
We omit the first term $\Omega_0$, which contains the ultraviolet divergence in infinite volume.
This representation reproduces the Casimir energy for $D=4$ and $\delta=2$, which is shown in Eqs.~(\ref{Casimir_antiperiodic}) and (\ref{Casimir_periodic}).

When we analytically derive the gap equation, the following relation of the Bessel function is useful:
\be{
\frac{d}{dx}\left(x^\nu K_\nu(ax)\right)
=
-ax^\nu K_{\nu-1}(ax).
}

Finally, we comment on anisotropic finite volume as a more special situation.
For example, we can consider finite volume for $L_1 \ll L_2 \neq \infty$.
Then, from Eq.~(\ref{Cas_ene_gene}), the finite-volume effects are dominated by contribution from the smaller $L_1$.
This situation is the same as competitions between finite volume and temperature, such as small volume at low temperature ($L \ll \beta$) and large volume at high temperature ($\beta \ll L$).

\section{Parity-doublet model with the six-point scalar vertex} \label{App_2}

In this Appendix, we check the effect of the six-point scalar vertex in the parity-doublet model.
This interaction was first introduced in Ref.~\cite{Motohiro:2015taa} to reproduce the incompressibility of nuclear matter.
The nuclear matter without this interaction was investigated in the early works~\cite{Hatsuda:1988mv, Zschiesche:2006zj, Sasaki:2010bp}.
Instead of Eq.~(\ref{pd_mes}), we set the following mesonic part of the Lagrangian,
\begin{eqnarray}
\mathcal{L}_\mathrm{Mirror}^\mathrm{mes} &=& \frac{1}{2}( \partial_\mu \sigma \partial^\mu \sigma) + \frac{1}{2}( \partial_\mu \vec{\pi} \partial^\mu \vec{\pi}) \nonumber\\
&& + \frac{\bar{\mu}^2 }{2} (\sigma^2 +\vec{\pi}^2) - \frac{\lambda}{4}  (\sigma^2 +\vec{\pi}^2)^2 + \frac{\lambda_6}{6}  (\sigma^2 +\vec{\pi}^2)^3 \nonumber\\
&& + \epsilon \sigma -\frac{1}{4} \omega_{\mu \nu} \omega^{\mu\nu} + \frac{m_\omega^2}{2} \omega_\mu \omega^\mu,
\end{eqnarray}
where $\lambda_6$ is the six-point coupling constant.
The numerical parameters based on Ref.~\cite{Motohiro:2015taa} are shown in Table~\ref{Tab_para_mir_lambda6}.\footnote{Notice that the parameters shown in the Erratum of Ref.~\cite{Motohiro:2015taa} include errors.
The correct parameters are those in the original article.
}

\begin{table}[t!]
\centering
\caption{Parameters of the parity-doublet model with the six-point scalar vertex~\cite{Motohiro:2015taa}, where $f_\pi=93 \mathrm{MeV}$, $m_\pi=140 \mathrm{MeV}$, and $\epsilon =m_\pi^2 f_\pi$.}
\begin{tabular}{cc}
\hline\hline
Parameters &  Values \\
\hline
$m_0$ (MeV) & $500$ \\
$g_1$ & $15.4$ \\
$g_2$ & $8.96$ \\
$g_\omega$ & $11.4$ \\
$\bar{\mu}$ (MeV) & $435$ \\
$\lambda$ & $40.5$ \\
$\lambda_6$ (MeV$^{-2}$)& $1.88 \times 10^{-3}$ \\
\hline\hline
\end{tabular}
\label{Tab_para_mir_lambda6}
\end{table}

\begin{figure}[t!]
    \begin{minipage}[t]{1.0\columnwidth}
        \begin{center}
            \includegraphics[clip, width=1.0\columnwidth]{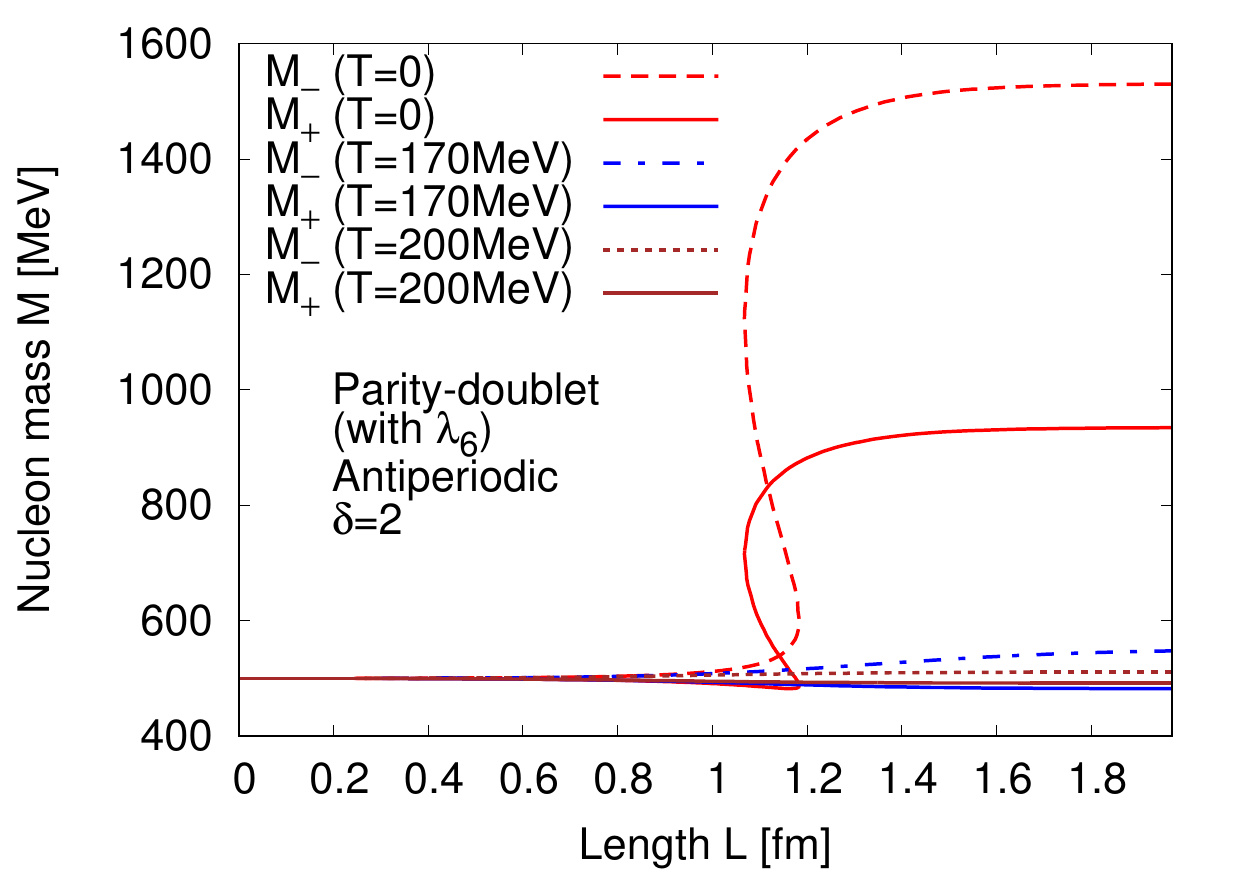}
        \end{center}
    \end{minipage}
    \begin{minipage}[t]{1.0\columnwidth}
        \begin{center}
            \includegraphics[clip, width=1.0\columnwidth]{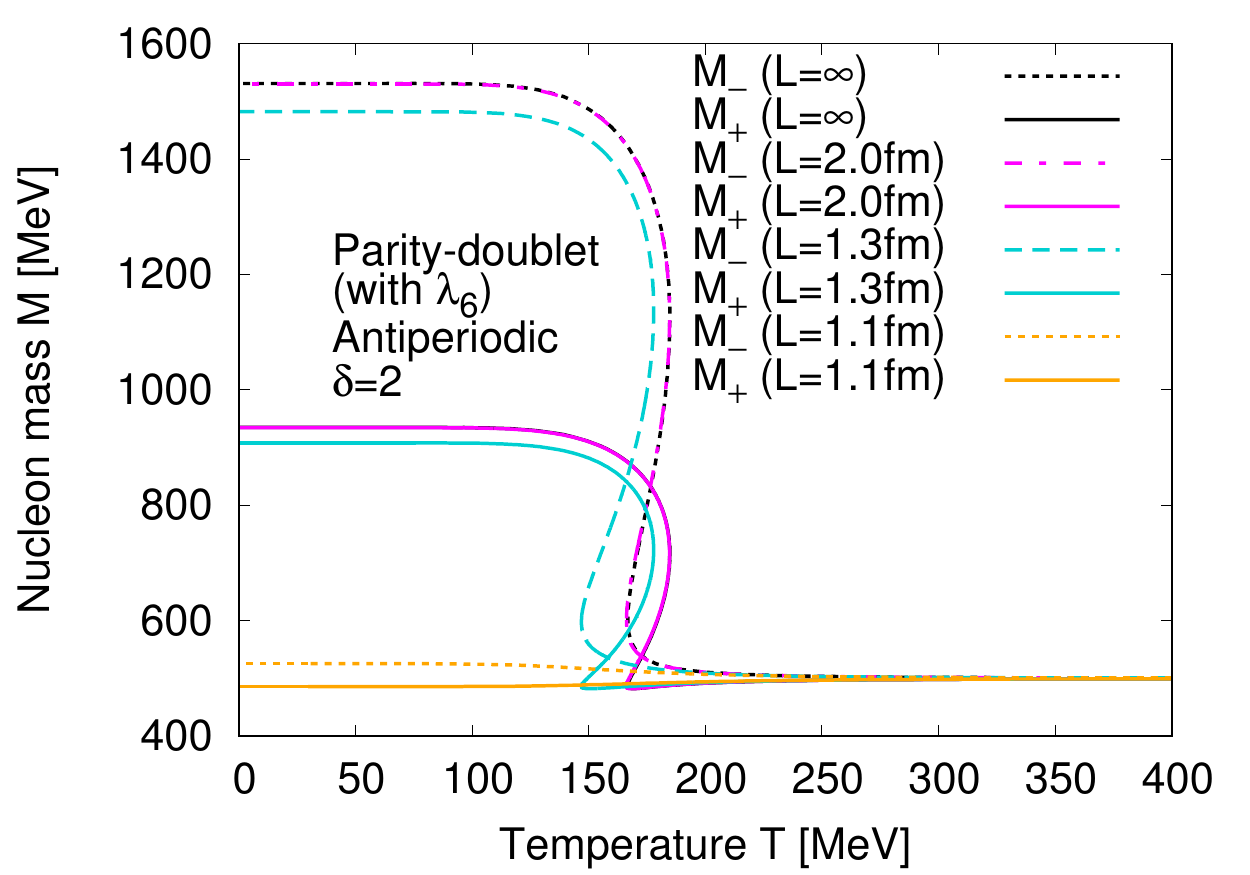}
        \end{center}
    \end{minipage}
    \caption{Nucleon masses with $\delta=2$ and antiperiodic boundary condition in the parity-doublet model {\it with the six-point scalar vertex}.
Upper: Finite-volume transition. Lower: Finite-temperature transition.}
    \label{apb_delta2_lambda6}
\end{figure}

\begin{figure}[t!]
    \begin{minipage}[t]{1.0\columnwidth}
        \begin{center}
            \includegraphics[clip, width=1.0\columnwidth]{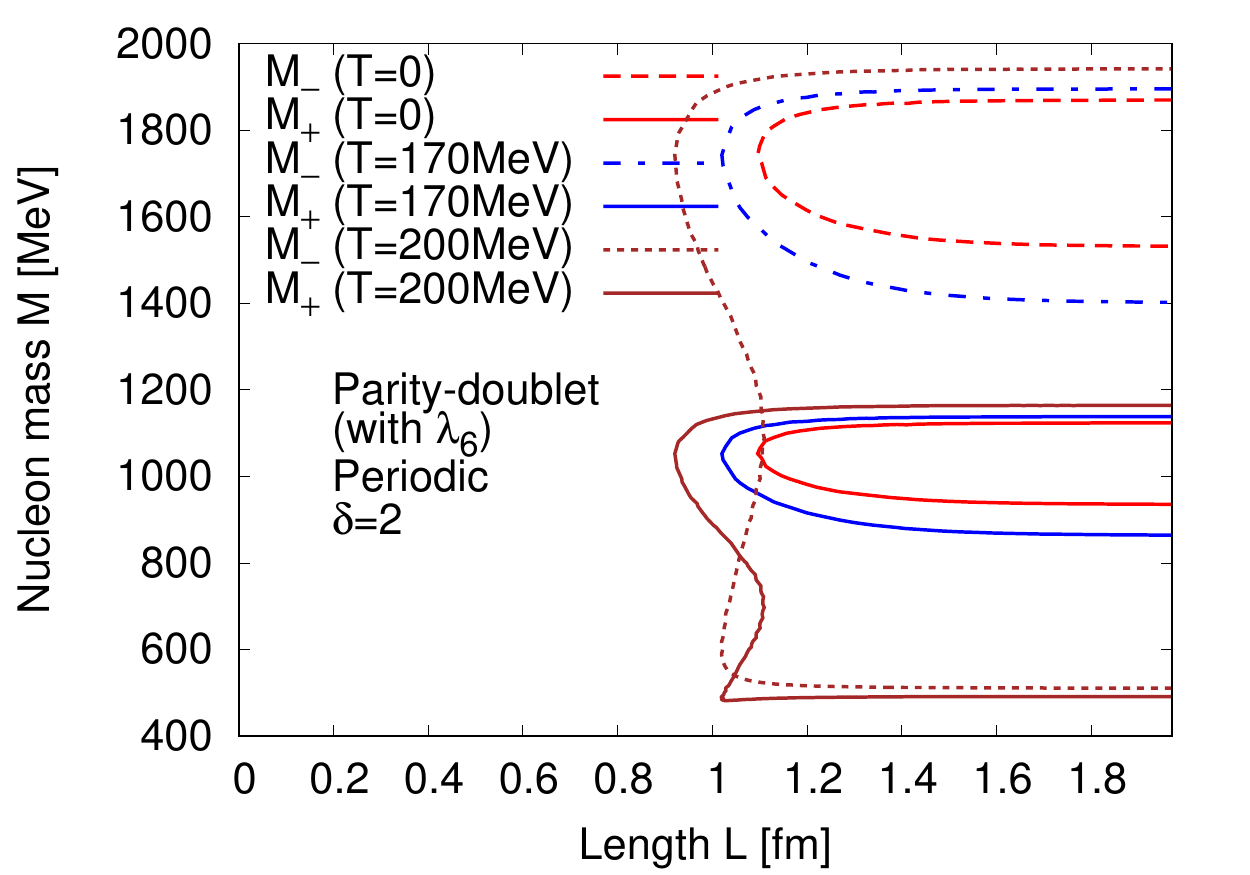}
        \end{center}
    \end{minipage}
    \caption{Finite-volume transition for nucleon masses with $\delta=2$ and periodic boundary condition in the parity-doublet model {\it with the six-point scalar vertex}.}
    \label{pb_LM_delta2_lambda6}
\end{figure}

In Fig.~\ref{apb_delta2_lambda6}, we show the phase transitions for the antiperiodic boundary condition.
In the setup with the six-point vertex, the orders of the finite-volume phase transition at $T=0$ and the thermal phase transition at $L=\infty$ are first order.
We find that, as $T$ increases, the finite-volume transition becomes a crossover.
Similarly, as $L$ decreases, the thermal phase transition becomes a crossover.

In Fig.~\ref{pb_LM_delta2_lambda6}, we show the results for the periodic boundary condition.
Notice that, in this figure, not only the minimum of the potential but also the maximum are shown as a solution to the gap equation.
Therefore, among the multiple solutions, the lower lines are favored.
For example, the lines starting from $M_+ \sim 940 \ \mathrm{MeV}$ and $M_- \sim 1500 \ \mathrm{MeV}$ at $T=0$ are favored.
In this case, we find behavior different from that of the results without the six-point vertex as shown in Fig.~\ref{pb_LM_delta2}.
In the small $L$ region, we find the disappearance of the solution for the nucleon masses (or the $\bar{\sigma}$ mean field). 
This is because the six-point vertex term has a minus sign in the thermodynamic potential, so the potential becomes unstable for a large value of $\bar{\sigma}$.
When $L$ is large enough, there is a (local) minimum of the thermodynamic potential, which is stabilized by the four-point scalar vertex term with a positive sign.
As $L$ decreases, the minimum is shifted to the larger $\bar{\sigma}$, and eventually, it becomes unstable by the six-point vertex.
Thus, in the small $L$ (or large $\bar{\sigma}$) region for the periodic boundary condition, this setup leads to the instability.
At least, we emphasize that, from Figs.~\ref{pb_LM_delta2} and \ref{pb_LM_delta2_lambda6}, the results in the large $L$ region are consistent within the parity-doublet model.
Such an instability by the six-point vertex could be improved by introducing higher-order terms.
In other words, the finite volume with a small $L$ is outside the scope of the parity-doublet model with the six-point scalar vertex because of its implicit UV cutoff.

\bibliography{nucleon}

\end{document}